\documentclass[pre, twocolumn, showpacs, superscriptaddress]{revtex4}
\usepackage{amsmath}
\usepackage{graphics}

\newcommand{\bra}{\left\langle}
\newcommand{\ket}{\right\rangle}
\newcommand{\kettr}{\right\rangle^{\rm tr}}
\newcommand{\pder}[2]{\frac{\partial #1}{\partial  #2}}
\newcommand{\pdert}[2]{\frac{\partial^2 #1}{\partial  #2^2}}
\newcommand{\der}[2]{\frac{\mathrm{d} #1}{\mathrm{d}  #2}}

\renewcommand{\d}{\mathrm{d}}
\newcommand{\e}{\mathrm{e}}

\newcommand{\ve}{\varepsilon}
\newcommand{\fp}{f^{\rm p}} 

\newcommand{\pc}{p_{\rm c}}
\newcommand{\vs}{\bar v} 

\newcommand{\vtr}[2]{\left( \begin{array}{c} #1 \\ #2 \end{array} \right)}
\newcommand{\mtx}[4]{\left( \begin{array}{cc} #1 & #2 \\ #3 & #4 \end{array} \right)}

\renewcommand{\Vec}[1]{\mbox{\boldmath $#1$}}
\newcommand{\sVec}[1]{\mbox{\scriptsize \boldmath $#1$}}

\begin{document}

\title{Energy dissipation and violation of the fluctuation-response relation in non-equilibrium Langevin systems}

\author{Takahiro Harada}
\email[Electronic address: ]{harada@chem.scphys.kyoto-u.ac.jp}
\affiliation{Department of Physics, Graduate School of Science, Kyoto University, Kyoto 606-8502, Japan}

\author{Shin-ichi Sasa}
\email[Electronic address: ]{sasa@jiro.c.u-tokyo.ac.jp}
\affiliation{Department of Pure and Applied Sciences, University of Tokyo, Komaba, Tokyo 153-8902, Japan}

\date{\today}

\begin{abstract}
The fluctuation-response relation is a fundamental relation that is applicable to systems near equilibrium. 
On the other hand, when a system is driven far from equilibrium, this relation is violated in general because the detailed-balance condition is not satisfied in nonequilibrium systems.
Even in this case, it has been found that for a class of Langevin equations, there exists an equality between the extent of violation of the fluctuation-response 
relation in the nonequilibrium steady state and the rate of energy dissipation from the system into the environment [T.~Harada and S.~-i.~Sasa, Phys.~Rev.~Lett. \textbf{95}, 130602 (2005)].
Since this equality involves only experimentally measurable quantities, it serves as a proposition to determine experimentally whether the system can be described by a Langevin equation.
Furthermore, the contribution of each degree of freedom to the rate of energy dissipation can be determined based on this equality.
In this paper, we present a comprehensive description on this equality, and provide a detailed derivation for various types of models including many-body systems, Brownian motor models, time-dependent systems, and systems with multiple heat reservoirs.
\end{abstract}


\maketitle


\section{Introduction} \label{s.intro}


In equilibrium statistical mechanics, the thermodynamic properties of a system in contact with a heat bath can be described in terms of the dynamical degrees of freedom $\Vec x$ instead of a complete set of variables of the system and heat bath $(\Vec x, \Gamma)$.
It is also widely accepted to describe a system in terms of effective variables $\Vec x$ along with an effective Hamiltonian $H(\Vec x)$ in which the contribution of the remaining variables $\Gamma$ are renormalized. By appropriately selecting effective variables $\Vec x$, we can study the universal features of a system \cite{Goldenfeld:1992}.


Similarly, in order to describe a system under a nonequilibrium condition, it may be reasonable to determine a closed description in terms of a part of the degrees of freedom $\Vec x$ instead of considering the entire set of variables $( \Vec x, \Gamma)$.
In this case, since we should consider the dynamics of the selected variables, such a reduction in the degrees of freedom requires the following strong assumption: (A1) {\it The typical time scale $\tau_{\sVec x}$ of the selected variables $\Vec x$ is considerably larger than the time scale $\tau_\Gamma$ of the remaining variables $\Gamma$.}
This assumption implies that there exists a time scale $\Delta t$ such that $\tau_\Gamma \ll \Delta t \ll \tau_{\sVec x} $.
When this condition is satisfied, the time evolution of the variables $\Vec x$ with a time interval of $\Delta t$ is described as a Markovian stochastic process; in particular, a Langevin equation is obtained in the limit $\Delta t/\tau_{\sVec x} \to 0$ \cite{vanKampen:1992, Gardiner:2004}.
The contribution of the eliminated degrees of freedom $\Gamma$ is in part renormalized into an effective Hamiltonian, and is in part decomposed into dissipation and noise terms of the Langevin equation characterized by a friction coefficient $\gamma$ and the noise intensity $M$, respectively.

Although condition (A1) is sufficient to describe the dynamical properties of the effective variables, it is insufficient to account for the thermodynamic properties of the system of $\Vec x$, particularly, the energetics.
In other words, condition (A1) alone does not guarantee that  the eliminated degrees of freedom can be regarded as a heat bath for the system of $\Vec x$.
To clarify this fact, let us consider a simple example \cite{Hayashi:2004}. Consider that a colloidal particle suspended in an aqueous solution of temperature $T$ is subjected to a periodic potential and constant driving force. With regard to the long-time behavior of the particle, we can obtain an effective description expressed as a Langevin equation with effective friction coefficient $\gamma$ and noise intensity $M$. However, in such a description, it is found that $M \neq \gamma T$, thereby implying that the second kind of the fluctuation-dissipation theorem cannot be applied naively in contrast to the equilibrium case (the Boltzmann constant is set to unity). This is because the eliminated variables, which include the short-time motion of the particle itself, are also affected by the driving force. In such a case,  it becomes difficult to identify the boundary between the system and the heat bath, and we cannot accurately determine the amount of energy transferred between $\Vec x$ and $\Gamma$.
In order to avoid this difficultly, we require another assumption: (A2) {\it The nonequilibrium condition imposed on the system of $\Vec x$ does not directly affect the eliminated degrees of freedom $\Gamma$}.
Based on assumptions (A1) and (A2), we can consider that the eliminated variables are quickly equilibrated at a temperature $T$ when no energy is transferred between $\Vec x$ and $\Gamma$.
When the system of $\Vec x$ exerts a force on the eliminated degrees of freedom, the reaction to the system of $\Vec x$ can be characterized with the linear-response properties of the eliminated variables, such as the friction coefficient $\gamma$.
In particular, when the system dynamics are described by a Langevin equation, assumption (A2) relates $M$ to $\gamma$ and $T$ according to the second kind of the fluctuation-dissipation theorem: $M = \gamma T$.
It should be noted that assumption (A2) corresponds to the local detailed-balance condition, which is regarded as a key property of stochastic processes for describing non-equilibrium steady states, since $M = \gamma T$ is derived by imposing this condition on a Langevin equation with $\gamma$ and $M$.


In systems under equilibrium conditions, the derivation of the Langevin equation from a classical mechanical system was formulated by employing the projection-operator method \cite{Zwanzig:1960, Mori:1965, Kawasaki:1973}.
However, for nonequilibrium conditions, there is no satisfactory theory that provides the foundation for the Langevin equation on the basis of a mechanical system. This is because it is difficult to treat persistent energy transfer from the degrees of freedom $\Vec x$ to the degrees of freedom $\Gamma$  mathematically.
Thus, the use of a Langevin equation to describe the dynamics of a nonequilibrium system in general is not justified, although a Langevin-type model is phenomenologically employed in many cases.


Based on this background, it is extremely important to {\it experimentally} validate such an effective description for nonequilibrium systems, i.e., validate assumptions (A1) and (A2).
For instance, consider the description of the above-mentioned system comprising a colloidal particle suspended in an aqueous solution.
We assume that an external force is exerted to drive the colloidal particle.
In this case, $\Vec x$ represents the position of the center of mass of the particle as $\Vec x = (x_0, x_1, x_2)$, and $\Gamma$ is defined such that $(\Vec x, \Gamma)$ represents all positions and conjugate momenta of atoms constituting the particle and the solution in an adiabatic container.
Static effects of the solvent can be incorporated in the effective Hamiltonian $H(\Vec x)$.
With these variables, it might be plausible to accept the assumptions (A1) and (A2).
However, even in such a simple case, their validity for a non-equilibrium state can be justified only with experimental confirmation.


This approach might be considered to be very strict.
However, when a system becomes more complicated, we must be very careful while accepting the validity of an effective description.
For example, in the past decade, many attempts have been made to describe the motion of a biological motor protein in terms of several effective variables using a Langevin-type stochastic model, termed Brownian motor model \cite{Julicher:1997, Reimann:2002}.
On the other hand, by considering the complex structure of the protein molecule, there is no definite reason to select a certain variable as an effective one from many internal degrees of freedom of the molecule.
Therefore, it is useful to establish a criterion to determine {\it experimentally} whether the degrees of freedom selected in a model satisfy assumptions (A1) and (A2). 


For this purpose, we need a proposition to validate assumptions (A1) and (A2) experimentally.
A proposition is favorable when it does not include any fitting parameters and involves quantities for explicit measurements.
Further, the best proposition for this purpose is the one that can be rigorously proved in a general class of Langevin-type models that satisfy assumptions (A1) and (A2). This is because the experimental examination of this proposition will allow us to validate assumptions (A1) and (A2) directly.


In order to formulate such a proposition, we determine certain experimentally measurable quantities.
In particular, we consider the nonequilibrium steady state of a system comprising several colloidal particles suspended in a solution under a nonequilibrium condition.
We select the spatial coordinates of the center of mass of the particles as the effective variables $\Vec x = (x_0, x_1, \cdots, x_{N-1})$. 
The basic statistical quantities are the steady current defined as
\begin{equation}
\vs_i \equiv \bra \dot x_i (t) \ket_0,
\label{e.vs}
\end{equation}
and the correlation function of velocity fluctuations
\begin{equation}
C_{ij}(t)\equiv \bra \left[ \dot x_i (t) -\vs_i \right] \left[ \dot x_j (0) - \vs_j \right] \ket_0,
\label{e.c}
\end{equation}
where $\bra \cdots \ket_0$ represents an ensemble average for the non-equilibrium steady state.


Evidently, $\vs_i$ and $C_{ij}(t)$ depend on system details such as an interaction potential between the particles.
In order to compare the experimental result with the computed value for a theoretical model, it is essential to tune the parameter values of the theoretical model.
Thus, the determination of only $\vs_i$ and $C_{ij}(t)$ is insufficient for validating the effective description. 


In order to obtain more information on the system, let us apply a small perturbation force $\ve \Vec f^\mathrm{p} (t) = \ve (\fp_0 (t), \fp_1(t), \cdots, \fp_{N-1}(t) )$ to it.
If its magnitude is sufficiently small ($\ve \ll 1$), we can expect that $\dot{\Vec x}$ will linearly respond to the perturbation as
\begin{equation}
\bra \dot x_i (t) \ket_\ve - \vs_i 
= \ve \sum_{j=0}^{N-1} \int_{-\infty}^t R_{ij} (t-s) \fp_j (s) \d s + O(\ve^2),
\label{e.r}
\end{equation}
in the limit $\ve \to 0$, where $\bra \cdots \ket_\ve$ represents the ensemble average in the presence of the perturbation force.
$R_{ij} (t)$ is termed the response function.

According to the fluctuation-dissipation theorem \cite{Kubo:1991}, if the system is in equilibrium, the response function carries the same 
information as the correlation function. This equivalence is expressed by the relation  
\begin{equation}
C_{ij} (t) = T R_{ij} (t) \qquad {\rm for} \quad  t > 0,
\label{e.FDT}
\end{equation}
which (termed ``the fluctuation-response relation'' hereafter) can be proved by assuming that the system satisfies the detailed-balance condition.
A noteworthy feature of the fluctuation-response relation is that it is closed only with experimentally measurable quantities.
Hence, the experimental determination of $C_{ij} (t)$ and $R_{ij} (t)$ enables us to verify whether the system satisfies the detailed-balance condition.


For the non-equilibrium steady state in which the detailed-balance condition is not satisfied, it is known that the fluctuation-response relation of Eq.~(\ref{e.FDT}) is violated \cite{Harada:2004, Hayashi:2004, Cugliandolo:1997a, Cugliandolo:1997}.
It should be noted that the response function defined in Eq.~(\ref{e.r}) characterizes the linear-response property of the nonequilibrium steady state and not that of the equilibrium state.
Therefore, the measurement of the response function provides information that differs from that of the correlation function.
However, since $R_{ij}(t)$ also depends on the system details, it is unsuitable for validating the effective description.


In such a case, if the violation can be expressed in a universal form, it is expected that this form may measure the ``distance'' of the nonequilibrium system from equilibrium.
Recently, it has been proved that a nonequilibrium Langevin model that satisfies (A1) and (A2), the extent of the violation is related to the rate of energy dissipation into the heat bath $\bra J \ket_0 $ by an equality \cite{Harada:2005c}.
According to this theory, the following equality holds provided the evolution of $\Vec x$ is determined by a Langevin equation: 
\begin{equation}
\bra J \ket_0 = \sum_{i=0}^{N-1} \gamma_i \left\{ \vs_i^2 + \int_{-\infty}^\infty \left[ \tilde C_{ii} (\omega) - 2 T \tilde R'_{ii} (\omega) \right] \frac{\d \omega}{2\pi} \right\},
\label{e.th}
\end{equation}
where $\gamma_i$ denotes the friction coefficient of $x_i(t)$; $\tilde C_{ij} (\omega)$ and $\tilde R_{ij} (\omega)$ represent the Fourier 
transforms of $C_{ij}(t)$ and $R_{ij} (t)$, respectively.
Similarly, the Fourier transform of an arbitrary function $A(t)$ is denoted by $\tilde A(\omega) \equiv \int_{-\infty}^\infty A(t) e^{\sqrt{-1} \omega t} \d t$; the prime denotes the real part.
In general, when the dynamics of $\Vec x(t)$ are overdamped, $\gamma_i^{-1} = \lim_{\omega \to \infty} \tilde R'_{ii} (\omega)$ holds.
It should be noted that the right-hand side of Eq.~(\ref{e.th}) represents the extent of violation of the fluctuation-response relation.


Next, we show that Eq. (\ref{e.th}) qualifies as the best proposition for experimental examination in order to investigate the validity of the effective description based on assumptions (A1) and (A2).
First, it is evident that Eq.~(\ref{e.th}) represents a closed relation among experimentally measurable quantities, without a fitting parameter.
Moreover, $\bra J \ket_0$ can be obtained by measuring the input energy, because in the nonequilibrium steady state, energy is externally injected at a constant rate and dissipated into the environment at the same rate.
Second, as shown in Ref.~\cite{Harada:2005c}, Eq.~(\ref{e.th}) holds for systems far from equilibrium, when the evolution of $\Vec x$
is described by a Langevin-type model.
The equality is independent of the other details of the model.
Therefore, it enables quantitative examination of the relevance of a Langevin-type model to the system under investigation.

Since a few simple examples were addressed in Ref.~\cite{Harada:2005c}, we provide a detailed description of the equality for several  Langevin models of physical interest.
In the following sections, we analyze many-body systems with and without inertia terms, stochastically or periodically driven systems, and systems in contact with multiple heat reservoirs.
We will show that it is possible to obtain a similar result for the relation between the dissipation rate and the extent of violation of the fluctuation-response relation irrespective of the model details. Further, we suggest a possible experimental study on this issue.


This paper is organized as follows.
In Sec.~\ref{s.many}, a Langevin model with many degrees of freedom is introduced as an example.
Then, Eq.~(\ref{e.th}) is proved for this model with and without the inertia terms, followed by several remarks.
In Sec.~\ref{s.ratchet}, generalized forms of Eq. (\ref{e.th}) are proved for  other cases, such as a model with a stochastically switching force, a model driven by a time-dependent external force, and a model that includes multiple heat reservoirs. 
Concluding remarks are provided in Sec.~\ref{s.conc} along with a suggestion of experimental studies related to this topic and future theoretical problems.
The proof of the fluctuation-dissipation theorem for Langevin systems under equilibrium is provided in Appendix \ref{a.FDT}.
In Appendix \ref{a.quick}, Eq.~(\ref{e.th}) is derived for the case of a single variable by using a path-integral argument.
Finally, in Appendix \ref{a.lemma}, the proof of a technical lemma used in the proof in Sec.~\ref{s.many} is provided.


\section{Many-body Langevin Systems} \label{s.many}

In this section, Eq.~(\ref{e.th}) is derived for a 
Langevin model with many variables. 
In particular, a model of colloidal suspension under non-equilibrium conditions  is studied, although the argument can be applied to various Langevin systems 
of many variables. The model and its energetic interpretation 
are explained in Sec.~\ref{ss.model}. Then, 
we present the mathematical proofs of  Eq.~(\ref{e.th})
for cases with and without the inertial terms in Secs.~\ref{ss.ud} and ~\ref{ss.od}, respectively.
In Sec.~\ref{ss.rem}, we comment on the result.

\subsection{Model} \label{ss.model}

We consider a three-dimensional system that comprises $n \equiv N/3$ spherical particles suspended in an aqueous solution. 
For this system, we adopt assumptions (A1) and (A2) as mentioned in Sec.~\ref{s.intro}, by considering the positions  and velocities of the center of mass of the particles as the effective variables. The three-dimensional position and velocity of the $\mu^\mathrm{th}$ particle are denoted by $\vec r_\mu \equiv ( r_\mu^0, r_\mu^1, r_\mu^2)$ and $\vec u_\mu \equiv (u_\mu^0, u_\mu^1, u_\mu^2)$, respectively, where $\mu = 0, 1, \cdots, n-1$.
Hereafter,  we collectively denote the positions and velocities of the particles as $\Vec{x} = (x_0, \cdots, x_{N-1}) \equiv (r_0^0, r_0^1, r_0^2, \cdots,  r_\mu^0, r_\mu^1, r_\mu^2, \cdots, r_{n-1}^0, r_{n-1}^1, r_{n-1}^2)$ and
$\Vec{v} = (v_0, \cdots, v_{N-1}) \equiv (u_0^0, u_0^1, u_0^2, \cdots, u_\mu^0, u_\mu^1, u_\mu^2, \cdots, u_{n-1}^0, u_{n-1}^1, u_{n-1}^2)$, respectively.

Based on assumption (A1), the motion of the particles is described by the  Langevin equations
\begin{eqnarray}
\dot x_i (t) &=& v_i (t) \label{e.model1} \\
m_i \dot v_i (t) &=& - \gamma_i v_i (t) + F_i (\Vec x (t)) + \xi_i (t) + \ve \fp_i (t), \label{e.model2}
\end{eqnarray}
where $i = 0, \cdots, N-1$. 
In this case, $m_i$ and $\gamma_i$ represent the mass and friction coefficient of the $\lfloor i/3 \rfloor^\mathrm{th}$ particle, respectively, where $\lfloor a \rfloor$ represents the largest integer that is not larger than $a$. 
Further, based on assumption (A2), $\xi_i (t)$ is the zero-mean white Gaussian noise that satisfies
\begin{equation}
\bra \xi_i(t)\xi_j(t') \ket=2\gamma_i T \delta_{ij} \delta (t - t').
\label{e.nint}
\end{equation}
The last term on the right-hand side of Eq.~(\ref{e.model2}) represents a probe force on the $i^\mathrm{th}$ coordinate with $0 \le \ve \ll 1$.
An initial condition is imposed at $t = t_0$, and we consider statistical quantities in the limit $t_0 \to -\infty$.


The second term $F_i (\Vec x)$ on the right-hand side of Eq.~(\ref{e.model2}) represents the force acting on the $i^\mathrm{th}$ coordinate.
For example, consider a system of colloidal particles trapped in an optical potential; a constant driving force $f \vec e_0$ is applied to them (see Ref.~\cite{Korda:2002}). This system may be described by selecting $F_i (\Vec x)$ as
\begin{eqnarray}
\lefteqn{F_i (\Vec x) = f \sum_{\mu = 0}^{n-1} \delta_{i, 3\mu} - \pder{}{x_i} \sum_{\mu = 0}^{n-1} U_\mu (\vec r_\mu) } \hspace{20mm} \nonumber \\
& & - \pder{}{x_i} \sum_{\mu = 0}^{n-1} \sum_{\nu = 0}^{n-1} \frac{V_{\mu \nu} (|\vec{r}_\mu - \vec{r}_\nu|)}{2},
\label{e.drive}
\end{eqnarray}
where $U_\mu (\vec r)$ represents an optical potential of the $\mu^\mathrm{th}$ particle and $V_{\mu \nu} (|\vec r|)$ is an interaction potential between the $\mu^\mathrm{th}$ and $\nu^\mathrm{th}$ particles.
Similarly, for a system of colloidal suspension under shear flow described by $\vec u (\vec r) = \hat \kappa \vec r$, where $\hat \kappa$ is a constant shear rate tensor, we can select $F_i(\Vec x)$ as
\begin{eqnarray}
\lefteqn{F_i (\Vec x) = \sum_{\mu = 0}^{n-1} \sum_{q = 0}^2 \sum_{q' = 0}^2 \delta_{i, (3\mu + q)} \kappa_{q q'} r_\mu^{q'} }\hspace{15mm} \nonumber \\
& & - \pder{}{x_i} \sum_{\mu = 0}^{n-1} \sum_{\nu = 0}^{n-1} \frac{V_{\mu \nu} (|\vec{r}_\mu - \vec{r}_\nu|)}{2},
\label{e.share}
\end{eqnarray}
For a detailed analysis on the behavior of this model, see Ref.~\cite{Fuchs:2005}.

Therefore, the model described by Eqs.~(\ref{e.model1}) and (\ref{e.model2}) can 
exhibit various phenomena of many-body Langevin systems; the following 
argument does not depend on the form of the selected force term $F_i (\Vec x)$.
However, it should be noted that the effect of hydrodynamic interaction between the particles is not included in Eqs.~(\ref{e.model1}) and (\ref{e.model2}).
Hence, the phenomena described by the model in these equations are rather ideal.
This might be justified when the mean distances between the particles are sufficiently large such that hydrodynamic correlation can be neglected.
In general, since there are several methods for including the effect of hydrodynamic interactions between particles \cite{Doi:1987}, this problem will be addressed elsewhere. 


Let us define the measurable quantities of this system.
Steady currents, velocity correlation functions, and response functions are already defined in Eqs.~(\ref{e.vs}), (\ref{e.c}), and (\ref{e.r}), respectively.
It is well known that when the system is in equilibrium, i.e., $f = 0$ in the case of Eq.~(\ref{e.drive}) and $\hat \kappa = 0$ in the case of Eq.~(\ref{e.share}),  the fluctuation-response relation described in Eq. (\ref{e.FDT}) holds. This will be demonstrated in Appendix \ref{a.FDT}. On the other hand, when the force terms contain nonconservative parts, the fluctuation-response relation is violated in steady states.

Next, we define the rate of energy dissipation according to K.~Sekimoto's argument \cite{Sekimoto:1997}. 
As discussed in Ref.~\cite{Sekimoto:1997}, it is natural to define the energy dissipated through the $i^\mathrm{th}$ coordinate during an infinitesimal interval $\Delta t$ as
\begin{equation}
J_i (t) \Delta t \equiv \int_t^{t+\Delta t} \left[ \gamma_i v_i (s) - \xi_i (s) \right] \circ \d x_i (s),
\label{e.diss}
\end{equation}
where the symbol $\circ$ denotes the multiplication in the sense of Stratonovich \cite{Gardiner:2004}.
The total rate of dissipation is the sum of $J_i (t)$'s as $J(t) = \sum_{i = 0}^{N-1} J_i (t)$.
Using the Stratonovich calculus, it is easy to show that this definition of the dissipation complies with the energy conservation law. For instance, in the case of  the force model represented by Eq.~(\ref{e.drive}), by summing Eq.~(\ref{e.diss}) over $i$ and setting $\ve = 0$, we obtain
\begin{widetext}
\begin{equation}
J(t) \Delta t = - \int_t^{t+\Delta t} \mathrm{d} \sum_{\mu = 0}^{n-1} \left[ \frac{m_{3\mu}}{2} |\vec u_\mu (s) |^2 + U_\mu (\vec r_\mu (s)) +  \sum_{\nu = 0}^{n-1} \frac{V_{\mu \nu} (|\vec r_\mu (s) - \vec r_\nu (s) |)}{2} \right] 
 + \sum_{\mu = 0}^{n-1} \left( f \vec e_0, \vec u_\mu (t) \right) \Delta t.
\label{e.cons}
\end{equation}
\end{widetext}
Since the first and second terms on the right-hand side of Eq.~(\ref{e.cons}) represent the change in the mechanical energy of the particles and the amount of energy input by the constant driving force, respectively, this equation expresses energy conservation by interpreting $J_i (t)$ defined in Eq.~(\ref{e.diss}) as the energy dissipated into the heat bath through the $i^\mathrm{th}$ degree of freedom.
Furthermore, the identification of energy dissipation in this manner was shown to agree with the second law of thermodynamics \cite{Sekimoto:1997b}.

\subsection{Proof:  Underdamped case} \label{ss.ud}

With this background, we prove the main result in Eq.~(\ref{e.th}).
First, let us express Eqs.~(\ref{e.model1}) and (\ref{e.model2}) in mathematical forms:
\begin{eqnarray}
\d x_i (t) &=& v_i (t) \d t \label{e.sanal1} \\
\d v_i (t) &=& \frac{- \gamma_i v_i (t) + F_i (\Vec x (t))}{m_i} \d t + \frac{\sqrt{2 \gamma_i T}}{m_i} \d W_i (t) \nonumber \\
& & + \frac{\ve \fp_i (t)}{m_i} \d t, \label{e.sanal2}
\end{eqnarray}
where $W_i (t)$ denotes a Wiener process \cite{Gardiner:2004}; $W_i (t)$ and $W_j (t)$ are assumed to be uncorrelated when $i \neq j$.
Using the It\^o formula, the time derivative of an arbitrary function $A(\Vec x(t), \Vec v(t))$ is calculated as
\begin{widetext}
\begin{eqnarray}
\d A(\Vec x(t), \Vec v(t)) &=& \sum_{i = 0}^{N-1}  v_i (t) \pder{}{x_i} A (\Vec x(t), \Vec v(t)) \d t \nonumber \\
& & + \sum_{i = 0}^{N-1} \left[ \frac{- \gamma_i v_i (t) + F_i (\Vec x(t))}{m_i} \pder{}{v_i} A (\Vec x(t), \Vec v(t)) + \frac{\gamma_i T}{m_i^2} \pdert{}{v_i} A (\Vec x(t), \Vec v(t)) \right] \d t \nonumber \\
& & + \sum_{i = 0}^{N-1} \pder{}{v_i} A (\Vec x(t), \Vec v(t)) \cdot \left[ \frac{\sqrt{2 \gamma_i T}}{m_i} \d W_i (t) + \frac{\ve}{m_i} \fp_i (t) \d t \right],
\label{e.ito}
\end{eqnarray}
\end{widetext}
where $\partial_{x_i} A (\Vec x (t), \Vec v (t) )$ represents $\partial_{x_i} A(\Vec x, \Vec v)$ evaluated at $(\Vec x, \Vec v) = (\Vec x(t), \Vec v(t))$.
A similar convention is used throughout this paper.
The symbol $\cdot$ denotes multiplication in the sense of It\^o \cite{Gardiner:2004}.
In conventional notation, Eq.~(\ref{e.ito}) can be rewritten as
\begin{eqnarray}
\lefteqn{\der{}{t} A(\Vec x(t), \Vec v(t)) = \Lambda A(\Vec x (t), \Vec v (t))} \hspace{5mm} \nonumber \\
& & + \sum_{i = 0}^{N-1} \pder{}{v_i} A(\Vec x(t), \Vec v(t)) \cdot \frac{\xi_i (t) + \ve \fp_i (t)}{m_i},
\label{e.Asde}
\end{eqnarray}
where
\begin{equation}
\Lambda \equiv \sum_{i=0}^{N-1} \left[ v_i \pder{}{x_i} + \frac{- \gamma_i v_i + F_i (\Vec x)}{m_i} \pder{}{v_i} + \frac{\gamma_i T}{m_i^2} \pdert{}{v_i} \right]
\label{e.bKramers}
\end{equation}
is the backward Kramers operator.
A solution of Eq.~(\ref{e.Asde}) can be expressed in the form
\begin{equation}
A(\Vec x(t), \Vec v(t)) = \mathcal{G}(t) A(\Vec x (t_0), \Vec v(t_0)),
\label{e.Ansatz}
\end{equation}
where the operator $\mathcal{G}(t)$ is independent of $A(\Vec x, \Vec v)$.
By substituting Eq.~(\ref{e.Ansatz}) into Eq.~(\ref{e.Asde}), we obtain a stochastic differential equation for $\mathcal{G}(t)$ as
\begin{equation}
\der{\mathcal{G}(t)}{t} = \mathcal{G}(t) \Lambda + \sum_{i = 0}^{N-1} \mathcal{G}(t) \pder{}{v_i} \cdot \frac{\xi_i (t) + \ve \fp_i (t)}{m_i},
\label{e.Gsde}
\end{equation}
with the initial condition, $\mathcal{G} (t_0) = 1$.
A formal solution of Eq.~(\ref{e.Gsde}) is expressed as
\begin{equation}
\mathcal{G} (t) = \e^{(t-t_0) \Lambda} + \int_{t_0}^t \sum_{i=0}^{N-1} \mathcal{G} (s) \pder{}{v_i} \e^{(t-s) \Lambda} \cdot \frac{\xi_i (s) + \ve \fp_i (s)}{m_i} \d s.
\label{e.Gformal}
\end{equation}
By identifying $A(\Vec x, \Vec v) = F_i (\Vec x)$ in Eq.~(\ref{e.Ansatz}) and using Eq.~(\ref{e.Gformal}), we obtain
\begin{eqnarray}
\lefteqn{F_i (\Vec x(t)) = \e^{(t-t_0)\Lambda} F_i (\Vec x(t_0))} \\
& & + \int_{t_0}^t \sum_{j = 0}^{N-1} \Phi_{ij} (t-s, \Vec x(s), \Vec v(s)) \cdot \left[ \xi_j (s) + \ve \fp_j (s) \right] \d s, \nonumber 
\label{e.Fformal}
\end{eqnarray}
where
\begin{equation}
\Phi_{ij} (t, \Vec x, \Vec v) \equiv \left\{ \begin{array}{ll}
\displaystyle \frac{1}{m_j} \pder{}{v_j} \e^{t \Lambda} F_i (\Vec x) &\quad \mathrm{for}~t >0 \\ 0 & \quad \mathrm{for}~t < 0
\end{array} \right. .
\end{equation}
By substituting Eq.~(\ref{e.Fformal}) into Eq.~(\ref{e.model2}) and averaging it with $\ve = 0$, it is shown that
\begin{equation}
\lim_{t_0 \to -\infty} \e^{(t-t_0) \Lambda} F_i (\Vec x(t_0)) = \gamma_i \vs_i,
\end{equation}
since the left-hand side does not depend on $t$.
Hereafter, the limit $t_0 \to -\infty$ will be considered.

The formal integration of Eq.~(\ref{e.model2}) yields
\begin{equation}
v_i(t) = \int_{-\infty}^t  H_i (t-s) \cdot \left[ F_i (\Vec x(s)) + \xi_i (s) +\ve \fp_i (s) \right] \d s,
\label{e.vformal}
\end{equation}
where 
\begin{equation}
H_i (t) \equiv \left\{ \begin{array}{ll} \displaystyle \frac{1}{m_i} \e^{-\gamma_i t / m_i } &\quad \mathrm{for}~t > 0 \\ 0 &\quad \mathrm{for}~t < 0 \end{array} \right. .
\end{equation}
By using Eq.~(\ref{e.Fformal}), we rewrite Eq.~(\ref{e.vformal}) as
\begin{eqnarray}
\lefteqn{v_i (t) - \vs_i = } \\
& & \int_{-\infty}^t \sum_{j = 0}^{N-1} K_{ij} \left(t-s, \Vec x(s), \Vec v(s)\right ) \cdot \left[ \xi_j (s) + \ve \fp_j (s) \right] \d s, \nonumber 
\label{e.FRE}
\end{eqnarray}
where
\begin{equation}
K_{ij} (t, \Vec x, \Vec v) \equiv  \int_0^\infty H_i (s) \Phi_{ij} (t-s, \Vec x, \Vec v) \d s + H_i (t) \delta_{ij}.
\end{equation}

Therefore, the average of Eq.~(\ref{e.FRE}) is expressed as
\begin{eqnarray}
\lefteqn{\bra v_i (t) \ket_\ve - \vs_i } \nonumber \\
&=& \ve \int_{-\infty}^t \sum_{j = 0}^{N-1} \bra K_{ij} \left( t-s, \Vec x(s), \Vec v(s) \right) \ket_0 \fp_j (s) \d s \nonumber \\
& &+ O(\ve^2).
\label{e.avFRE}
\end{eqnarray}
Since Eq.~(\ref{e.avFRE}) holds irrespective of the functional form of $\fp_i (t)$, by comparing Eqs.~(\ref{e.r}) and (\ref{e.avFRE}), we obtain
\begin{equation}
R_{ij} (t-s) = \bra K_{ij} \left( t - s, \Vec x(s) , \Vec v(s) \right) \ket_0.
\label{e.RK}
\end{equation}

Next, Eq.~(\ref{e.diss}) is rewritten as
\begin{equation}
J_i (t) \Delta t = \int_t^{t+\Delta t} \left[ \gamma_i v_i (s)^2 \d s - \sqrt{2 \gamma_i T} v_i (s) \circ \d W_i (s) \right].
\label{e.diss2}
\end{equation}
Now, we set $\ve = 0$. By using the lemma described in Appendix \ref{a.lemma}, Eqs.~(\ref{e.FRE}) and (\ref{e.RK}) lead to 
\begin{equation}
\bra v_i(s) \circ \d W_i (s) \ket_0 = \lim_{s \to 0+} \sqrt{\frac{\gamma_i T}{2}} R_{ii}(s) \d s
\end{equation}
By definition, since $R_{ij} (t) = 0$ for $t < 0$ 
\begin{equation}
\lim_{t \to 0+} R_{ii} (t) = 2 \int_{-\infty}^\infty \tilde R'_{ii} (\omega) \frac{\d \omega}{2\pi}.
\end{equation}
From
\begin{eqnarray}
\bra v_i (t)^2 \ket_0 &=& \vs_i^2 + C_{ii} (0) \nonumber \\
&=& \vs_i^2 + \int_{-\infty}^\infty \tilde C_{ii} (\omega) \frac{\d \omega}{2\pi},
\end{eqnarray}
we finally obtain
\begin{equation}
\bra J_i \ket_0 = \gamma_i \left\{ \vs_i^2 + \int_{-\infty}^\infty \left[ \tilde C_{ii} (\omega) - 2T \tilde R'_{ii} (\omega) \right] \frac{\d \omega}{2 \pi} \right\}.
\label{e.singleth}
\end{equation}
Since the total rate of dissipation $J(t)$ is the sum of the rates of dissipation through the $i^\mathrm{th}$ degree of freedom $J_i (t)$, Eq.~(\ref{e.th}) is immediately obtained.

\subsection{Proof: Overdamped case} \label{ss.od}

In Eqs.~(\ref{e.model1}) and (\ref{e.model2}), when $m_i=0$, the Langevin equation takes an overdamped form:
\begin{equation}
\gamma_i \dot x_i (t) 
= F_i \left(\Vec x(t) \right) + \xi_i (t) + \ve \fp_i (t).
\label{e.od}
\end{equation}
For this model, the proof of equalities in Eq.~(\ref{e.singleth}) requires a special treatment because  $C_{ii}(0)$ and $R_{ii}(0)$ are divergent in this case.
In the following, we prove Eq.~(\ref{e.singleth}) by considering this singularity.
In this case, we interpret the correlation function as $C_{ij} (t) \equiv \bra [\dot x_i (t) - \vs_i ] \circ [\dot x_j (0) - \vs_j ] \ket_0$.

First, by using the It\^o formula, the time evolution of an arbitrary function $A( \Vec x(t))$ is expressed as
\begin{equation}
\der{}{t} A\left( \Vec x(t) \right) = \Lambda A \left( \Vec x(t) \right) + \sum_{i = 0}^{N-1} \pder{}{x_i} A(\Vec x(t)) \cdot \frac{\xi_i (t) + \ve \fp_i (t)}{\gamma_i},
\label{e.itod}
\end{equation}
where 
\begin{equation}
\Lambda \equiv \sum_{i = 0}^{N-1} \left[ \frac{F_i \left(\Vec x(t) \right)}{\gamma_i} \pder{}{x_i} + \frac{T}{\gamma_i} \pdert{}{x_i} \right]
\label{e.FP}
\end{equation}
is the backward Fokker-Planck operator.
In order to solve Eq.~(\ref{e.itod}), we introduce an operator $\mathcal{G} (t)$ as
\begin{equation}
A(\Vec x(t)) = \mathcal{G}(t) A(\Vec x(t_0)),
\label{e.gdefod}
\end{equation}
where $\mathcal{G}(t)$ is independent of $A(\Vec x)$.
By substituting Eq.~(\ref{e.gdefod}) into Eq.~(\ref{e.itod}), we obtain an equation for $\mathcal{G}(t)$ as
\begin{equation}
\der{\mathcal{G}(t)}{t} = \mathcal{G}(t) \Lambda + \sum_{i = 0}^{N-1} \mathcal{G}(t) \pder{}{x_i} \cdot \frac{\xi_i (t) + \ve \fp_i (t)}{\gamma_i}.
\label{e.geqod}
\end{equation}
A formal solution of Eq.~(\ref{e.geqod}) is given by
\begin{equation}
\mathcal{G}(t) = \e^{(t-t_0) \Lambda} + \int_{t_0}^t \sum_{i=0}^{N-1} \mathcal{G}(s) \pder{}{x_i} \e^{(t-s)\Lambda} \cdot \frac{\xi_i (s) + \ve \fp_i (s)}{\gamma_i} \d s.
\label{e.gformalod}
\end{equation}
By setting $A(\Vec x) = F_i (\Vec x)$, we immediately obtain
\begin{eqnarray}
\lefteqn{F_i (\Vec x(t)) = \e^{(t-t_0) \Lambda} F_i (\Vec x(t_0)) }\nonumber \\
& &+ \int_{t_0}^t \sum_{j=0}^{N-1} \Phi_{ij} (t-s, \Vec x(s)) \cdot \left[ \xi_j (s) + \ve \fp_j (s) \right] \d s,
\label{e.fformalod}
\end{eqnarray}
where
\begin{equation}
\Phi_{ij} (t, \Vec x) \equiv \left\{ \begin{array}{ll}
\displaystyle \frac{1}{\gamma_j} \pder{}{x_j} \e^{t\Lambda}F_i (\Vec x) & \mathrm{for}~t > 0 \\ 0 & \mathrm{for}~t < 0 .
\end{array} \right. 
\end{equation}
By substituting Eq.~(\ref{e.fformalod}) into Eq.~(\ref{e.od}) and averaging it with $\ve = 0$, it is found that
\begin{equation}
\lim_{t_0 \to -\infty} \e^{(t-t_0) \Lambda} F_i (\Vec x(t_0)) = \gamma_i \vs_i.
\end{equation}
Hereafter, we consider the limit $t_0 \to -\infty$. The substitution of Eq.~(\ref{e.fformalod}) into Eq.~(\ref{e.od}) leads to the equation
\begin{equation}
\dot x_i (t) - \vs_i = \int_{-\infty}^t \sum_{j=0}^{N-1} K_{ij} (t-s, \Vec x(s)) \cdot \left[ \xi_j (s) + \ve \fp_j (s) \right] \d s,
\label{e.FREod}
\end{equation}
where
\begin{equation}
K_{ij} (t, \Vec x) \equiv \left\{ \begin{array}{ll}
\displaystyle \frac{1}{\gamma_i} \left[ \Phi_{ij} (t, \Vec x) + \delta_{ij} \delta(t) \right] & \mathrm{for}~t > 0 \\ 0 & \mathrm{for}~t < 0.
\end{array} \right. 
\label{e.Kod}
\end{equation}
By comparing Eq.~(\ref{e.r}) with the average of Eq.~(\ref{e.FREod}), we find
\begin{equation}
R_{ij} (t-s) = \bra K_{ij} \left(t-s, \Vec x(s) \right) \ket_0.
\label{e.RKod}
\end{equation}

Henceforth, we discretize the time as $t^k = k \Delta t$ with an interval $\Delta t$ in order to clarify our argument mathematically.
By setting $\ve = 0$, the discretized form of the Langevin equation (\ref{e.od}) becomes
\begin{eqnarray}
\gamma_i \Delta x_i^k &=& \frac{ F_i (\Vec x^{k+1}) + F_i (\Vec x^k) }{2} \Delta t + \sqrt{2 \gamma_i T} \Delta W_i^k \nonumber \\
& & + O\left(\Delta t^{3/2}\right)
\label{e.discLangevin}
\end{eqnarray}
where $\Vec x^k = \{ x_i^k \} \equiv \{ x_i (t^k) \}$, $\Delta x_i^k \equiv x_i^{k+1} - x_i^k$, and $\Delta W_i^k \equiv W_i (t^{k+1}) - W_i (t^k)$ (see  Appendix \ref{a.FDT}).
Similarly, Eq.~(\ref{e.diss}) is discretized as
\begin{equation}
J_i (t^k) \Delta t = \frac{F_i (\Vec x^{k+1}) + F_i (\Vec x^k)}{2} \Delta x_i^k + O\left( \Delta t^2 \right),
\label{e.discdiss}
\end{equation}
where Eq.~(\ref{e.od}) and the definition of the symbol $\circ$ have been used.
By combining Eqs.~(\ref{e.discLangevin}) and (\ref{e.discdiss}), a straightforward calculation yields
\begin{eqnarray}
\bra J_i (t^k) \ket_0 &=& \gamma_i \vs_i^2 + \gamma_i \bra \left( \frac{\Delta x_i^k}{\Delta t} - \vs_i \right)^2 \ket_0 - \frac{2T}{\Delta t} \nonumber \\
& & - \sqrt{\frac{2T}{\gamma_i}} \frac{\bra  \left[ F_i (\Vec x^{k+1}) + F_i (\Vec x^k) \right] \Delta W_i^k \ket_0}{2 \Delta t} \nonumber \\
& & + O\left( \Delta t^{1/2} \right).
\label{e.Jdisc}
\end{eqnarray}
For the limit $\Delta t \to 0$, the second and third terms on the right-hand side of Eq.~(\ref{e.Jdisc}) can be transformed as
\begin{eqnarray}
\lefteqn{\lim_{\Delta t \to 0} \left[ \gamma_i \bra \left( \frac{\Delta x_i^k}{\Delta t} - \vs_i \right)^2 \ket_0 - \frac{2T}{\Delta t} \right] } \hspace{20mm}\nonumber \\
&=& \int_{-\infty}^\infty \left[ \gamma_i \tilde C_{ii} (\omega) - 2T \right] \frac{\d \omega}{2\pi}.
\label{e.Jdisc1}
\end{eqnarray}

Next, the discretized form of Eq.~(\ref{e.fformalod}) with $\ve = 0$ becomes
\begin{equation}
F_i (\Vec x^k) = \gamma_i \vs_i + \sum_{l = 1}^\infty \sum_{j = 0}^{N-1} \Phi_{ij} (t^l, \Vec x^{k-l}) \sqrt{2 \gamma_j T} \Delta W_j^{k-l}.
\label{e.Fdisc}
\end{equation}
Hence, the fourth term on the right-hand side of Eq.~(\ref{e.Jdisc}) is calculated as
\begin{equation}
\sqrt{\frac{2T}{\gamma_i}} \frac{\bra \left[ F_i (\Vec x^{k+1}) + F_i (\Vec x^k) \right] \Delta W_i^k \ket_0}{2\Delta t} = T \bra \Phi_{ii}(\Delta t, \Vec x^k) \ket_0,
\end{equation}
where the relation $\bra \Delta W_i^k \Delta W_j^l \ket_0 = \delta_{ij} \delta_{kl} \Delta t$ is used.
According to Fourier's theorem, 
\begin{eqnarray}
\lefteqn{\lim_{\Delta t \to 0+} \frac{\bra \Phi_{ii}(\Delta t, \Vec x^k) \ket_0 + \bra \Phi_{ii}(- \Delta t, \Vec x^k) \ket_0}{2}}\hspace{20mm} \nonumber \\
&=&  \int_{-\infty}^{\infty} \left[ \gamma_i \tilde R'_{ii} (\omega) - 1 \right] \frac{\d \omega}{2\pi},
\label{e.Fourth}
\end{eqnarray}
where Eqs.~(\ref{e.Kod}) and (\ref{e.RKod}) are used.
By combining Eqs.~(\ref{e.Jdisc}), (\ref{e.Jdisc1}), and (\ref{e.Fourth}), we finally obtain the following expression with the limit $\Delta t \to 0+$:
\begin{equation}
\bra J_i \ket_0 = \gamma_i \left\{ \vs_i^2 + \int_{-\infty}^\infty \left[ \tilde C_{ii} (\omega) - 2T \tilde R'_{ii} (\omega) \right] \frac{\d \omega}{2\pi} \right\}.
\label{e.singlethod}
\end{equation}
It should be noted that the integral on the right-hand side of Eq.~(\ref{e.singlethod}) is convergent in the limit $\Delta t \to 0+$.
Since the total rate of dissipation $J(t)$ is the sum of the dissipation rate through each degree of freedom $J_i (t)$, Eq.~(\ref{e.th}) is immediately obtained.

\subsection{Remarks} \label{ss.rem}

We present several comments on the proof in the previous subsections.
First, the final result is  independent of the selected force term $F_i(\Vec x)$.
In particular, since no smallness of the driving forces is assumed in the derivation, Eqs.~(\ref{e.singleth}) and (\ref{e.singlethod}) hold even when the system is far from equilibrium.
Their right-hand sides represent the extent of violation of the fluctuation-response relation. These equalities imply that the rate of energy dissipation through the $i^\mathrm{th}$ degree of freedom is directly related to the violation of the fluctuation-response relation for this degree of freedom.

Second, in the overdamped case, we obtain
\begin{equation}
\gamma_i^{-1} = \lim_{\omega \to \infty} \tilde R'_{ii} (\omega),
\label{e.gammaR}
\end{equation}
from Eq.~(\ref{e.RKod}) by using $\lim_{\omega \to \infty}\tilde \Phi_{ij} (\omega, \Vec x) =  0$. 
It should be noted that the inertial effect cannot be observed in standard experiments on colloidal systems.
Thus, all the quantities on the right-hand side of Eq.~(\ref{e.th}) can be directly measured experimentally.

Third, Eq.~(\ref{e.singleth}) can be rewritten in a more compact form in the underdamped case.
By using Fourier's theorem and Eq.~(\ref{e.RK}), we can calculate $\int_{-\infty}^\infty \tilde R'_{ii} (\omega) \d \omega/ 2\pi = [ R_{ii}(0+) + R_{ii}(0-) ]/2 = 1/(2m_i) $. Therefore, Eq.~(\ref{e.singleth}) can be expressed as
\begin{equation}
\bra J_i \ket_0 =\frac{\gamma_i}{m_i} \left[ m_i \bra v_i(t)^2 \ket_0 - T \right].
\label{e.kin}
\end{equation}
Thus, the rate of dissipation through the $i^\mathrm{th}$ degree of freedom can be expressed as the deviation of the kinetic temperature of this degree of freedom from the temperature of the heat bath.
In the overdamped case, we cannot give such an interpretation since the kinetic energy cannot be defined.
Moreover, for experimental use, Eq.~(\ref{e.singleth}) is more convenient than Eq.~(\ref{e.kin}), because in many experiments, accurate determination of the kinetic temperature requires an extremely fine time resolution.

Fourth, we demonstrate that the result presented in the previous subsections can be generalized further.
For example, the following quantity
\begin{eqnarray}
\lefteqn{I_{ij} (t) \equiv }\hspace{0mm} \\
&& \frac{1}{2} \bra v_i(t)\circ\left[ \gamma_j v_j (0) - \xi_j (0) \right] + v_i (0)\circ\left[ \gamma_j v_j (t) - \xi_j (t) \right] \ket_0 \nonumber
\label{e.gen}
\end{eqnarray}
can be rewritten in terms of the fluctuation-response relation violation as
\begin{eqnarray}
\lefteqn{I_{ij} (t) =} \\
& & \gamma_j \left\{ \vs_i \vs_j+ \int_{-\infty}^\infty \left[ \tilde C'_{ij} (\omega) - 2 T \tilde R'_{ij} (\omega) \right] e^{-\sqrt{-1} \omega t} \frac{\d \omega}{2 \pi} \right \}. \nonumber
\label{e.genth}
\end{eqnarray}
For the underdamped case, Eq.~(\ref{e.genth}) can be obtained by substituting Eq.~(\ref{e.FRE}) into Eq.~(\ref{e.gen}) and using the lemma in Appendix \ref{a.lemma} (see Appendix \ref{a.quick} for the overdamped case).
Since $I_{ii} (0) = \bra J_i \ket_0$, Eq.~(\ref{e.genth}) is regarded as a generalization of Eq.~(\ref{e.singleth}).
In addition, the diagonal elements of Eq.~(\ref{e.gen}) are force-velocity correlation functions expressed as
\begin{equation}
I_{ii} (t) =  \frac{1}{2}\bra v_i(t)\circ F_i (\Vec x(0)) + v_i(0)\circ F_i (\Vec x(t)) \ket_0,
\label{e.fvc}
\end{equation}
which is immediately obtained from Eqs.~(\ref{e.od}) and (\ref{e.gen}) with $\ve = 0$ in the overdamped case.
In the underdamped case, Eq.~(\ref{e.fvc}) follows from Eqs.~(\ref{e.model2}) and (\ref{e.gen}) and the fact that $\bra v_i (t) \circ \dot v_i (0) \ket_0 = - \bra v_i (0) \circ \dot v_i (t) \ket_0$.
In general, the physical significance of the off-diagonal elements of the violation has not yet determined.

Finally, we discuss a relation between Eq.~(\ref{e.th}) and linear response theory \cite{Kubo:1991, Groot:1984}.
In this theory, the power loss that is proportional to the square of the driving force is discussed in terms of the response function.
It should be noted that the response function considered in this theory is defined only at equilibrium. 
Therefore, this response function denoted by $R_{ij}^\mathrm{eq} (t)$ characterizes the linear response from the equilibrium state.
For example, we consider a force model $F_i (\Vec x) = f_i - \partial_{x_i} U(\Vec x)$, where $f_i$ is a constant driving force and $U(\Vec x)$ is a potential.
When the driving forces are sufficiently small, we can calculate the dissipation rate (linear power loss) from Eqs.~(\ref{e.model2}) and (\ref{e.diss}) and the definition of $R_{ij}^\mathrm{eq} (t)$ as
\begin{equation}
\bra J \ket_0 = \sum_{i =0}^{N-1} \sum_{j =0}^{N-1} \tilde R^\mathrm{eq}_{ij} (0) f_i f_j + O(f^2),
\label{e.linear}
\end{equation}
which is in accordance with  linear-response theory \cite{Kubo:1991, Groot:1984}.
On the other hand, our result in Eq.~(\ref{e.th}) is valid independent of the magnitude of the driving forces (It should be noted that $\tilde R_{ij} (\omega)$ in Eq.~(\ref{e.th}) differs from $\tilde R_{ij}^\mathrm{eq}(\omega)$ when $f_i \neq 0$).
Hence, Eq.~(\ref{e.th}) should agree with Eq.~(\ref{e.linear}) when the driving forces are sufficiently small.
Thus, it might be interesting to demonstrate directly the equivalence between these two expressions.


\section{Other examples} \label{s.ratchet}

In this section, it is demonstrated that the violation of the fluctuation-response relation is related to the energy dissipation for the other types of Langevin models.
We consider the following models: a model with stochastically switching potentials, a system driven by a temporally periodic force, and a system in contact with multiple heat reservoirs. These models were originally introduced phenomenologically in order to describe particular non-equilibrium phenomena without microscopic foundations. In the following subsections, we present a method by which Eq.~(\ref{e.th}) can be extended to each case; this will allow us to examine the relevance of each model to describe a certain phenomenon.

\subsection{Stochastically driven system} \label{ss.flash} 

First, a model with stochastically switching potentials is considered. For simplicity, we consider a model with one spatial degree of freedom.
Let $x$ be the position of a particle in one-dimensional space. We assume that this particle has an internal degree of freedom denoted by $\sigma$.
Moreover, let this particle be exerted a potential force $F_\sigma \equiv - \partial_x U_\sigma (x)$ depending on $\sigma$. 
Therefore, the model is expressed as
\begin{eqnarray}
\dot x (t) &=& v(t) \label{e.flash1} \\
m \dot v(t) &=& - \gamma v(t) + F_{\sigma(t)} (x(t)) + \xi(t) + \ve \fp (t), \label{e.flash2}
\end{eqnarray}
where $m$ and $\gamma$ denote the mass and friction coefficient of the particle, respectively, and $\xi(t)$ represents the zero-mean white Gaussian noise that satisfies
\begin{equation}
\bra \xi(t)\xi(t') \ket=2\gamma T\delta(t-t').
\end{equation}
As mentioned earlier, $\ve \fp(t)$ is a probe force with a sufficiently small $\ve$.
We assume that $\sigma (t)$ is a Poisson process in $\{0, 1 \}$. The transition rates from state 0 to state 1 and vice versa are denoted by $\Omega_{10} (x)$ and $\Omega_{01} (x)$, respectively; they can depend on the position $x$ of the particle.
The following analysis can be extended to the case that involves two or more internal states.
This type of model was originally suggested as the model of a motor protein; it was termed \textit{flashing ratchet} \cite{Julicher:1997, Reimann:2002, Okada:1999}.

For this model, the rate of energy dissipation $J(t)$ is defined as 
\begin{equation}
J(t) \Delta t \equiv \int_t^{t + \Delta t} \left[ \gamma v(s) - \xi(s) \right] \circ \d x(s),
\label{e.diss1}
\end{equation}
where $\Delta t$ is an infinitesimal time interval. It has been shown that the following equality holds \cite{Sekimoto:1997, Parrondo:1998}:
\begin{eqnarray}
J(t) \Delta t &=& -\int_{t}^{t+ \Delta t} \d \left[ \frac{m}{2} v(s)^2 + U_{\sigma(t)} (x(s)) \right] \\
& & + \sum_{j} \left[ U_{\sigma(\hat \tau^j + 0)} (x(\hat \tau^j)) - U_{\sigma(\hat \tau^j - 0)} (x(\hat \tau^j)) \right], \nonumber
\label{e.flashcons}
\end{eqnarray}
where $\ve = 0$ and $\hat \tau^j$ for $j=1,2,\cdots$ denotes the time at which the transition of the internal state $\sigma(t)$ occurs.
The summation of the second term on the right-hand side is over $j$ that satisfies  $t \le \hat \tau_j \le  t + \Delta t$. 
Since this term can be regarded as an energy gain accompanied with state transitions, Eq.~(\ref{e.flashcons}) can be interpreted as the energy conservation law in the case of this model.

Here, we show that the rate of energy dissipation $J(t)$ can be expressed in terms of the violation of the fluctuation-response relation as
\begin{equation}
\bra J \ket_0 = \gamma \left \{ \vs^2 + \int_{-\infty}^\infty \left[ \tilde C(\omega) - 2T \tilde R'(\omega) \right] \frac{\d \omega}{2\pi} \right\},
\label{e.flashth}
\end{equation}
where the definition of the velocity correlation function $C(t)$ and the response function $R(t)$ are similar to those in Eqs.~(\ref{e.c}) and (\ref{e.r}), respectively.
Eq.~(\ref{e.flashth}) is identical to Eq.~(\ref{e.th}) with a single spatial degree of freedom.

Now, we prove Eq.~(\ref{e.flashth}).
As in the previous case, we study the time evolution of an arbitrary function $A_{\sigma(t)} (x(t), v(t))$.
First, we fix a trajectory of the particle $\{ x(t), v(t) \}$ and a history of transitions represented by $\hat \tau^j$ for $j=1,2,\cdots$.
For this history, we select a small time interval $\Delta t$ in which 
at most one transition can occur.
Let $\hat Z_{\sigma'\sigma}(t;x(t)) \Delta t =1$ when a transition from a state $\sigma$ to another state $\sigma'$ occurs in the 
the interval $(t, t + \Delta t)$; otherwise $\hat Z_{\sigma'\sigma}(t;x(t)) \Delta t =0$.
It should be noted that the expectation value of $\hat Z_{\sigma'\sigma}(t;x(t))$ over the ensemble of transition histories for a fixed value of $x(t)$ is equal to the transition rate $\Omega_{\sigma'\sigma}(x(t))$. 

Then, the time evolution of $A_{\sigma(t)} (x(t), v(t))$ during $\Delta t$ is written as
\begin{eqnarray}
\lefteqn{ A_{\sigma(t + \Delta t)}(x(t+\Delta t), v(t+\Delta t)) 
- A_{\sigma(t)}(x(t),v(t)) 
} \hspace{10mm}\nonumber \\
&=& A_{\sigma(t + \Delta t)}(x(t), v(t)) - A_{\sigma(t)}(x(t),v(t))  \nonumber \\
& & + A_{\sigma(t)}(x(t+\Delta t), v(t+\Delta t)) \nonumber \\
& &- A_{\sigma(t)}(x(t),v(t)) +
O( \Delta t^{3/2}).
\label{e.Aexp}
\end{eqnarray}
The first term on the right-hand side can be expressed as
\begin{eqnarray}
\lefteqn{
A_{\sigma(t + \Delta t)}(x(t), v(t)) - A_{\sigma(t)}(x(t),v(t)) =} \hspace{0mm} \nonumber \\
& & \delta_{\sigma(t) 0} \left[ A_1 (x(t), v(t)) - A_0 (x(t), v(t)) \right] \hat Z_{10} (t; x(t)) \Delta t \nonumber \\
& & + \delta_{\sigma(t) 1} \left[ A_0 (x(t), v(t)) - A_0 (x(t), v(t)) \right] \hat Z_{01}(t; x(t)) \Delta t \nonumber \\
& & +O( \Delta t^{3/2}).
\label{e.state}
\end{eqnarray}
On the other hand, by using the It\^o formula, the second term on the right-hand side of Eq.~(\ref{e.Aexp})
is expressed as
\begin{widetext}
\begin{eqnarray}
\lefteqn{
A_{\sigma(t)}(x(t+\Delta t), v(t+\Delta t )) - A_{\sigma(t)}(x(t),v(t)) = \sum_{\sigma' \in \{0, 1\}} \delta_{\sigma(t) \sigma'} \Lambda_{\sigma'} A_{\sigma'} (x(t), v(t)) 
\Delta t
} \hspace{30mm} \nonumber \\
& & + \sum_{\sigma' \in \{0, 1\}} \delta_{\sigma(t) \sigma'} 
\pder{}{v} A_{\sigma'} (x(t), v(t)) 
\cdot \frac{\sqrt{2 \gamma T} \Delta W(t) + \ve \fp (t) \Delta t}{m}
+O( \Delta t^{3/2}),
\label{e.flashito}
\end{eqnarray}
\end{widetext}
where $\Lambda_\sigma$ denotes the backward Kramers operator corresponding to each state:
\begin{equation}
\Lambda_\sigma \equiv v \pder{}{x} + \frac{-\gamma v + F_\sigma(x)}{m} \pder{}{v} + \frac{\gamma T}{m^2} \pdert{}{v}.
\end{equation}
Hence, by combining Eqs.~(\ref{e.Aexp}), (\ref{e.state}), and (\ref{e.flashito}) and taking the limit $\Delta t \to 0$, we obtain a general expression of the time derivative of $A_{\sigma(t)} (x(t), v(t))$ as
\begin{widetext}
\begin{eqnarray}
\der{}{t} A_{\sigma(t)}(x(t), v(t)) &=&\left( \delta_{\sigma(t) 0}~\delta_{\sigma(t) 1} \right) \mtx{\Lambda_0 - \Omega_{10} (x(t))}{\Omega_{10}(x(t))}{\Omega_{01}(x(t))}{\Lambda_1 - \Omega_{01}(x(t))} \vtr{A_0 (x(t), v(t))}{A_1 (x(t), v(t))} \nonumber \\
& &+ \left( \delta_{\sigma(t) 0}~\delta_{\sigma(t) 1} \right) \frac{\xi(t) + \ve \fp(t)}{m} \pder{}{v} \cdot  \vtr{A_0 (x(t), v(t))}{A_1 (x(t), v(t))} \nonumber \\
& &+ \left( \delta_{\sigma(t) 0}~\delta_{\sigma(t) 1} \right) \cdot \mtx{- \zeta_{10} (t; x(t))}{\zeta_{10} (t; x(t))}{\zeta_{01} (t; x(t))}{-\zeta_{01} (t; x(t))} \vtr{A_0 (x(t), v(t))}{A_1 (x(t), v(t))},
\label{e.Aevo2}
\end{eqnarray}
\end{widetext}
where $\zeta_{\sigma' \sigma} (t; x(t)) \equiv \hat Z_{\sigma' \sigma} (t; x(t)) - \Omega_{\sigma' \sigma} (x(t))$.
The symbol $\cdot$ in the last line of Eq.~(\ref{e.Aevo2}) implies that $\zeta_{\sigma' \sigma} (t; x(t))$ is statistically independent of $\sigma(t)$ (It\^o-type definition).

Next, we introduce an operator $\mathcal{G}(t)$ such that
\begin{equation}
A_{\sigma(t)} (x(t), v(t)) = \mathcal{G}(t) A_{\sigma(t_0)} (x(t_0), v(t_0))
\label{e.Gflash}
\end{equation}
By substituting Eq.~(\ref{e.Gflash}) into Eq.~(\ref{e.Aevo2}), we obtain an equation for $\mathcal{G}(t)$
\begin{eqnarray}
 \dot{ \mathcal{G}}(t) &=& \mathcal{G}(t) \Lambda + \mathcal{G} (t)  \pder{}{v} \cdot \frac{\xi(t) + \ve \fp(t)}{m} \nonumber \\
 & & + \mathcal{G} (t) \cdot \zeta(t; x),
\label{e.Gevo}
\end{eqnarray}
where
\begin{equation}
\Lambda \equiv \left( \delta_{\sigma 0}~\delta_{\sigma 1} \right) \mtx{\Lambda_0 - \Omega_{10} (x)}{\Omega_{10}(x)}{\Omega_{01}(x)}{\Lambda_1 - \Omega_{01}(x)} \vtr{1}{1}
\end{equation}
and
\begin{equation}
\zeta(t; x) \equiv \left( \delta_{\sigma 0}~\delta_{\sigma 1} \right) \mtx{- \zeta_{10} (t; x)}{\zeta_{10} (t; x)}{\zeta_{01} (t; x)}{-\zeta_{01} (t; x)} \vtr{1}{1}.
\end{equation}
The initial condition is $\mathcal{G} (t_0) = 1$.
A formal solution of Eq.~(\ref{e.Gevo}) is 
\begin{eqnarray}
\mathcal{G} (t) &=& \e^{(t-t_0)\Lambda} + \int_{t_0}^t \mathcal{G}(s) \pder{}{v} \e^{(t-s) \Lambda} \cdot \frac{\xi(s)  + \ve \fp(s)}{m} \d s \nonumber \\
& &  + \int_{t_0}^t \mathcal{G}(s) \e^{(t-s) \Lambda} \cdot \zeta(s; x) \d s.
\end{eqnarray}

Therefore, the force $F_{\sigma(t)} (x(t))$ is expressed as
\begin{eqnarray}
\lefteqn{F_{\sigma(t)} (x(t)) = \left. \e^{(t-t_0)\Lambda} F_{\sigma(t_0)}(x(t_0)) \right|_{v=v(t_0)} }  \nonumber \\
 & & + \int_{t_0}^t \Phi_{\sigma(s)}(t-s, x(s), v(s)) \cdot \left[ \xi(s) + \ve \fp(s) \right] \d s \nonumber \\
& & + \int_{t_0}^t \left. \e^{(t-s)\Lambda} F_{\sigma(s)} (x(s)) \right|_{v=v(s)} \cdot \zeta(s; x(s)) \d s,
\label{e.flashF}
\end{eqnarray}
where
\begin{equation}
\Phi_\sigma (t, x, v) \equiv \left\{ \begin{array}{ll}
\displaystyle \frac{1}{m} \pder{}{v} \e^{t \Lambda} F_\sigma (x) &\quad \mathrm{for}~t > 0 \\ 0 &\quad \mathrm{for}~t < 0
\end{array} \right. .
\end{equation}
Since $\bra \zeta(s, x(s)) \ket_0 = 0$,
\begin{equation}
\lim_{t_0 \to -\infty} \left. \e^{(t-t_0)\Lambda} F_{\sigma(t_0)}(x(t_0)) \right|_{v=v(t_0)} = \gamma \vs.
\end{equation}
Henceforth, we consider the limit $t_0 \to -\infty$.

Next, Eq.~(\ref{e.flash2}) is formally solved as
\begin{equation}
v(t) = \int_{-\infty}^t  H (t-s) \cdot \left[ F_{\sigma(s)} (x(s)) + \xi (s) +\ve \fp_i (s) \right] \d s,
\label{e.vformalf}
\end{equation}
where 
\begin{equation}
H(t) \equiv \left\{ \begin{array}{ll}
\displaystyle \frac{1}{m} \e^{-\gamma t / m} & \quad \mathrm{for}~t > 0 \\ 0 & \quad \mathrm{for}~t < 0
\end{array} \right. .
\end{equation}
By substituting Eq.~(\ref{e.flashF}) into Eq.~(\ref{e.vformalf}), we obtain
\begin{eqnarray}
\lefteqn{v(t) - \vs = \int_{-\infty}^t K_{\sigma(s)} (t-s, x(s), v(s)) \cdot \left[ \xi(s) +\ve \fp(s) \right] \d s } \hspace{5mm} \nonumber \\
& & + \int_{-\infty}^t H(t-s) \int_{-\infty}^s \left. \e^{(s-s') \Lambda} F_{\sigma(s')} (x(s')) \right|_{v=v(s')} \nonumber \\
& &  \quad \quad \cdot \zeta(s'; x(s')) \d s' \d s,
\label{e.flashFRE}
\end{eqnarray}
where
\begin{equation}
K_{\sigma} (t, x, v) \equiv \int_0^\infty H(s) \Phi_\sigma (t-s, x, v) \d s + H(t).
\end{equation}
The average of Eq.~(\ref{e.flashFRE}) and a comparison with the definition of the response function results in
\begin{equation}
R(t-s) = \bra K_{\sigma(s)} (t-s, x(s), v(s)) \ket_0.
\label{e.flashRK}
\end{equation}
The right-hand side is a function of $t-s$ due to the time translational symmetry of the steady state.

Since Eq.~(\ref{e.diss1}) can be rewritten as
\begin{equation}
J(t) \Delta t = \int_t^{t+\Delta t} \left[ \gamma v (s)^2 \d s - \sqrt{2 \gamma T} v (s) \circ \d W (s) \right],
\label{e.dissm}
\end{equation}
we obtain Eq.~(\ref{e.singleth}) based on an argument similar to that in Sec.~\ref{ss.ud} and by using Eqs.~(\ref{e.flashFRE}) and (\ref{e.flashRK}).

\subsection{Time-dependent system} 

We consider a case in which the driving force is time dependent.
For simplicity, only a system with a single spatial degree of freedom is considered, although the analysis presented here can be generalized to a multi-dimensional case.
The model is expressed as
\begin{eqnarray}
\dot x (t) &=& v(t) \label{e.peri1} \\
m \dot v(t) &=& - \gamma v(t) + F(x(t), t) + \xi(t) + \ve \fp (t), \label{e.peri2}
\end{eqnarray}
where the notations are the same as those in Eqs.~(\ref{e.flash1}) and (\ref{e.flash2}).
The second term on the right-hand side of Eq.~(\ref{e.peri2}) $F(x, t)$ represents a time-dependent force.
For example, we might assume that $F(x, t)$ consists of conservative and non-conservative parts as $F(x, t) = - \partial_x U(x) + f(t)$, where $f(t)$ is a time-dependent driving force, although the final result is independent of this assumption.
This model can be regarded as the model of macroionic current in the presence of an AC electric field; it was also studied in the context of a Brownian ratchet \cite{Magnasco:1993}.

The rate of energy dissipation is defined according to Eq.~(\ref{e.diss1}).
In this case, the law of energy conservation is expressed as
\begin{eqnarray}
J(t) \Delta t &=& - \int_t^{t+ \Delta t} \d \left[ \frac{m}{2} v(s)^2 + U(x(s)) \right] \nonumber \\
& &+ \int_t^{t+\Delta t} f(s) \circ \d x(s).
\label{e.pericons}
\end{eqnarray}
Since the system does not possess time-translational invariance in the presence of the time-dependent driving force, we define the velocity correlation function as
\begin{equation}
C(\Delta, t) \equiv \bra \left[ v(t + \Delta) - \vs (t + \Delta) \right] \left[ v(t) - \vs (t) \right] \ket_0,
\label{e.peric}
\end{equation}
where $\vs(t) \equiv \bra v(t) \ket_0$ is the ensemble-averaged velocity at time $t$.
The response function in this case is defined as
\begin{equation}
\bra v(t) \ket_\ve - \vs (t) = \ve \int_{-\infty}^t R(t-s, s) \fp (s) \d s + O(\ve^2).
\label{e.perir}
\end{equation}

With this background, we present the equality between the rate of dissipation and the violation of the fluctuation-response relation:
\begin{equation}
\bra J(t) \ket_0 = \gamma \left\{ \vs (t)^2 + \int_{-\infty}^\infty \left[ \tilde C(\omega, t) - 2 T \tilde R'(\omega, t) \right] \frac{\d \omega}{2 \pi} \right\},
\label{e.perith}
\end{equation}
where $\tilde A(\omega, t) \equiv \int_{-\infty}^\infty A(\Delta, t) e^{i \omega \Delta} \d \Delta$ for an arbitrary function, $A(\Delta, t)$.
Thus, the result can be generalized for systems without  time-translational invariance.

We now derive Eq.~(\ref{e.perith}).
First, we introduce a new variable $\theta$ and rewrite Eqs.~(\ref{e.peri1}) and (\ref{e.peri2}) in an autonomous form as
\begin{eqnarray}
\dot x (t) &=& v(t), \label{e.perim1} \\
m \dot v(t) &=& - \gamma v(t) + F(x(t), \theta(t)) + \xi(t) + \ve \fp (t), \label{e.perim2} \\
\dot \theta (t) &=& 1 \label{e.perim3},
\end{eqnarray}
where $\theta(t_0) = t_0$.
By using the It\^o formula, the time evolution of an arbitrary function, $A(x(t), v(t), \theta(t))$, is obtained as
\begin{eqnarray}
\lefteqn{\der{}{t} A(x(t), v(t), \theta(t)) = \Lambda A(x(t), v(t), \theta(t))} \hspace{10mm} \nonumber \\
& & + \pder{}{v} A(x(t), v(t), \theta(t)) \cdot \frac{\xi(t) + \ve \fp(t)}{m},
\label{e.perito}
\end{eqnarray}
where
\begin{equation}
\Lambda \equiv v \pder{}{x} + \frac{-\gamma v + F(x, \theta)}{m} \pder{}{v} + \frac{\gamma T}{m^2} \pdert{}{v} + \pder{}{\theta}.
\label{e.periKra}
\end{equation}
We introduce an operator that is independent of $A$ such that
\begin{equation}
A(x(t), v(t), \theta(t)) = \mathcal{G} (t) A( x(t_0), v(t_0), \theta(t_0) ).
\label{e.perigdef}
\end{equation}
By substituting Eq.~(\ref{e.perigdef}) into Eq.~(\ref{e.perito}), we obtain a stochastic differential equation for $\mathcal{G}(t)$ as
\begin{equation}
\dot{\mathcal{G}} (t) = \mathcal{G} (t) \Lambda + \frac{\xi(t) + \ve \fp(t)}{m} \cdot \mathcal{G} (t) \pder{}{v},
\label{e.geq}
\end{equation}
where the initial condition is $\mathcal{G} (t_0) = 1$. A formal solution of Eq.~(\ref{e.geq}) is 
\begin{equation}
\mathcal{G} (t) = \e^{(t-t_0) \Lambda} + \int_{t_0}^t \mathcal{G}(s) \pder{}{v} e^{(t-s)\Lambda} \cdot \frac{\xi(s) + \ve \fp(s)}{m} \d s.
\label{e.formalperig}
\end{equation}
Therefore, $F(x(t), \theta(t))$ can be expressed as
\begin{eqnarray}
\lefteqn{ F(x(t), \theta(t)) = \left. \mathcal{G} (t) F(x(t_0), \theta(t_0)) \right|_{v=v(t_0)}} \\
&=& \left. \e^{(t-t_0)\Lambda} F(x(t_0), \theta(t_0)) \right|_{v=v(t_0)} \nonumber \\
& & + \int_{t_0}^t \Phi(t-s, x(s), v(s), \theta(s)) \cdot \left[ \xi(s) + \ve \fp(s) \right] \d s, \nonumber
\label{e.periformalf}
\end{eqnarray}
where
\begin{equation}
\Phi (t, x, v, \theta) \equiv \left\{ \begin{array}{ll}
\displaystyle \frac{1}{m} \pder{}{v} \e^{t \Lambda} F(x, \theta) &\quad \mathrm{for}~t > 0 \\ 0 &\quad \mathrm{for}~t < 0 
\end{array} \right. .
\label{e.periPhi}
\end{equation}

By taking the limit $t_0 \to -\infty$, the first term on the right-hand side of Eq.~(\ref{e.periformalf}) converges to a function of $t$ only:
\begin{equation}
\lim_{t_0 \to -\infty} \left. \e^{(t-t_0)\Lambda} F(x(t_0), \theta(t_0)) \right|_{v=v(t_0)} = \bar{F} (t).
\label{e.flim}
\end{equation}
We therefore derive
\begin{equation}
\gamma \vs (t) = \bar{F}(t).
\end{equation}

A formal solution of Eq.~(\ref{e.peri2}) is given as
\begin{equation}
v(t) = \int_{-\infty}^t  H (t-s) \cdot \left[ F (x(s), s) + \xi (s) +\ve \fp_i (s) \right] \d s,
\label{e.vformalp}
\end{equation}
where
\begin{equation}
H(t) \equiv \left\{ \begin{array}{ll}
\displaystyle \frac{1}{m} \e^{-\gamma t / m} &\quad \mathrm{for}~t > 0 \\ 0 &\quad \mathrm{for}~t < 0
\end{array} \right. .
\end{equation}
By substituting Eq.~(\ref{e.periformalf}) into Eq.~(\ref{e.vformalp}), we obtain
\begin{eqnarray}
\lefteqn{v(t) - \vs (t) = }\\
& & \int_{-\infty}^t K(t-s, x(s), v(s), \theta(s)) \cdot \left[ \xi(s) + \ve \fp(s) \right] \d s, \nonumber 
\label{e.periFRE}
\end{eqnarray}
where
\begin{equation}
K(t, x, v, \theta) \equiv \int_0^\infty H(s) \Phi(t-s, x, v, \theta) \d s + H(t).
\label{e.periK}
\end{equation}
The average of Eq.~(\ref{e.periFRE}) yields
\begin{equation}
R(t -s, s) = \bra K(t -s , x(s), v(s), \theta(s)) \ket_0.
\label{e.periRK}
\end{equation}
It should be noted that the $s$-dependence of the right-hand side of Eq.~(\ref{e.periRK}) is retained after taking the average; this can be confirmed from Eqs.~(\ref{e.perigdef}) and (\ref{e.formalperig}).

Since the definition of $J(t)$ in Eq.~(\ref{e.diss1}) can be rewritten as Eq.~(\ref{e.dissm}), we can obtain Eq.~(\ref{e.perith}) based on an argument similar to that in Sec.~\ref{ss.ud} and by using Eqs.~(\ref{e.periFRE}) and (\ref{e.periRK}).

\subsection{Multiple heat reservoirs} 

Finally, we address systems with multiple heat reservoirs by considering two cases.
The first case involves a heat bath with a spatially inhomogeneous temperature profile.
Such a model can be considered as a model of thermophoresis; it was first analyzed by M.~B\"uttiker and R.~Landauer \cite{Buttiker:1987, Landauer:1988}. 
In particular, we investigate the model represented by Eqs.~(\ref{e.model1}) and (\ref{e.model2}).
Let the spatial profile of the temperature be $T(\vec r)$.
In this case, the noise intensity $2\gamma_i T$ in Eq.~(\ref{e.nint}) is
replaced with 
$2 \gamma_i T\left( \vec r_\mu (t) \right)$ for $\mu = \lfloor i/3 \rfloor$.
In order to avoid any ambiguity due to multiplicative noise,
we assume that the model can be represented in the underdamped form, i.e., $m_i \not =0$ .

The definitions of the measurable quantities and the rate of energy dissipation are the same as Eqs.~(\ref{e.vs}), (\ref{e.c}), (\ref{e.r}), and (\ref{e.diss}).
For this case, the following equality is derived:
\begin{equation}
\bra J_i \ket_0 = \gamma_i \left\{ \vs_i^2 + \int_{-\infty}^\infty \left[ \tilde C_{ii} (\omega) - 2 \bar{T}_i \tilde R'_{ii} (\omega) \right] \frac{\d \omega}{2\pi} \right\},
\label{e.butth}
\end{equation}
where $\bar{T}_i \equiv \bra T \left(\vec r_\mu (\cdot) \right) \ket_0$ for $\mu = \lfloor i/3 \rfloor$ is the steady temperature averaged using a steady distribution with respect to $\Vec x$. Since the proof of Eq.~(\ref{e.butth}) is almost similar to that in Sec.~\ref{ss.ud}, we have not mentioned it here.

In the second case, each degree of freedom in a system is in contact with a different heat bath of a different temperature.
We reinvestigate the same underdamped model described by Eq.~(\ref{e.model1}) and (\ref{e.model2}); however, in this case, the temperature depends on the index of the degrees of freedom. Therefore, the variance of $\xi_i (t)$ is considered as $2 \gamma_i T_i$.
The definitions of the measurable quantities are identical to ones described by Eqs.~(\ref{e.vs}), (\ref{e.c}), and (\ref{e.r}). The definition of dissipation rates is the same as Eq.~(\ref{e.diss}).

For this model, we can prove the equality
\begin{equation}
\bra J_i \ket_0 = \gamma_i \left\{ \vs_i^2 + \int_{-\infty}^\infty \left[ \tilde C_{ii} (\omega) - 2 T_i \tilde R'_{ii} (\omega) \right] \frac{\d \omega}{2\pi} \right\}.
\label{e.condth}
\end{equation}
by replacing $T$ with $T_i$ in the proof given in Sec.~\ref{ss.ud}.

The above argument can be applied to the problem of heat conduction.
For instance, let us consider the one-dimensional lattice heat conduction.
We assume $x_i (t)$ to be the one-dimensional position of the $i^\mathrm{th}$ particle, and $v_i (t)$ to be its velocity.
The force term is selected as
\begin{equation}
F_i ( \Vec x) = - a (x_i - i \ell)
- b  (x_i - i \ell)^3+c (x_{i+1} - 2x_i+x_{i-1}) ,
\end{equation}
where $\ell$ denotes the lattice constant and $a$, $b$, and $c$ are constants.
We set $x_{-1} \equiv x_0$ and $x_N \equiv x_{N-1}$.
Sites at the both ends of the chain are assumed to be connected to heat baths 
of different temperatures as $T_0\ge T_{N-1}$, while the other sites are not connected to a heat bath: $\gamma_i = 0$ for $i \neq 0, N-1$.
Evidently, Eq.~(\ref{e.condth}) holds for this model. Further, $- \bra J_0 \ket_0 \ge 0$ represents the heat transferred from the high-temperature heat bath, and $\bra J_{N-1} \ket_0 \ge 0$ represents the heat dissipated into the low-temperature heat bath. Due to the energy balance in the system, $\bra J_0 \ket_0 + \bra J_{N-1} \ket_0 = 0$ holds. Therefore, $- \bra J_0 \ket_0 (= \bra J_{N-1} \ket_0)$ represents the heat flux through the system. It should be noted that 
$\bra J_i \ket_0 = 0$ for $i \neq 0, N-1$ in this case.
Thus, the heat flux through the system is explicitly related to the violation of the fluctuation-response relation at the end of the chain.
On the other hand, the relation between the heat flux in the system and the violation of the fluctuation-response relation inside (bulk) the system has not yet been determined.


\section{Concluding Remarks} \label{s.conc}
\subsection{Conclusion}

In this paper, we presented several results with regard to the relationship between the rate of energy dissipation and the violation of the fluctuation-response relation for various types of nonequilibrium Langevin models.
The most important feature of these results is that they enable the determination of the rate of energy dissipation based only on experimentally measurable quantities and without detailed knowledge on the system.
Hence, our results provide a proposition that can be experimentally verified. The experimental verification of the equality, when possible, ensures that the system is in fact a Langevin-type system, i.e., assumptions (A1) and (A2) are acceptable.
If the equality cannot be established experimentally, it implies the existence of other slow degrees of freedom that were not considered.
Hence, the equality presented in this paper serves as a ``check sum.''

The present result is also suitable for practical use.
If it has been already established that the system in concern is well described with a Langevin model, Eqs.~(\ref{e.singleth}) and (\ref{e.singlethod}) will provide a measure of the contribution of each degree of freedom to energy dissipation.
An advantage of our result is that we do not require the detailed knowledge on a system to determine the rate of dissipation.
This enables the determination of the relative importance of each degree of freedom in a complicated system from the viewpoint of energetics.

\subsection{Suggestion of experiments}

In order to demonstrate the above mentioned concepts, we suggest a possible experiment on a motor protein.
Y. Okada \textit{et al.}~reported that a motor protein termed KIF1A, a single-headed kinesin superfamily protein, can be modeled as a flashing ratchet model \cite{Okada:1999, Okada:2003}. This is because the microtubule exhibits a quasi one-dimensional periodic structure on which a KIF1A molecule moves processively and KIF1A has two internal states, strong and weak binding states, according to the chemical state of the nucleotide hydrolyzed in the molecule. Okada \textit{et al.}~explained the results of single-molecule experiments using a flashing ratchet model by adopting several fitting parameters \cite{Okada:2003}. However, the relevance of these parameters has not yet been experimentally confirmed, because of certain difficulties in experimental techniques.

If the argument that the KIF1A molecule can be described as a flashing ratchet is valid, Eq. (\ref{e.flashth}) should hold for this molecule according to the result in Sec.~\ref{ss.flash}.
As mentioned above, Eq.~(\ref{e.flashth}) can be verified without specifying the model parameters such as the profile of periodic potentials.
The right-hand side of Eq.~(\ref{e.flashth}) might be determined by employing the present techniques of the single-molecule experiment.
On the other hand, the rate of chemical free energy consumption by the motor molecule can be estimated by means of biochemical techniques.
If these quantities are in agreement, the relevance of a Langevin-type model to this molecule is quantitatively ensured. In other words, only the center of mass is the slow variable for this molecule, and it contributes to energy dissipation.
However, if the right-hand side of Eq.~(\ref{e.flashth}) is less than the rate of chemical energy input, it implies the existence of more degrees of freedom that should be considered and that the flashing ratchet model is inappropriate.

Moreover, the experimental determination of $\bra J_i \ket_0$ for the center of mass of the protein using Eq.~(\ref{e.flashth}) reveals the amount of chemical energy input that is converted into the translational motion of the motor molecule.
Since the question ``how much chemical energy is converted into mechanical energy?'' is one of the most important problems regarding a motor protein, such an experimental study will serve to answer it.

\subsection{Future perspectives}

Finally, we present future theoretical problems for consideration.
First, although our argument began with the Langevin equations, it should be possible to derive the same result by beginning with a microscopic mechanical model that satisfies the fundamental assumptions (A1) and (A2). This will not only provide another perspective of the problem but will also help to generalize the framework of the theory.
We remark that a simple case has been analyzed quite recently \cite{Teramoto:2005}.

Further generalizations of Eq.~(\ref{e.th}) for cases that are not considered in the present paper might be possible.
For example, it might be interesting to consider a case with a finite-time correlation of noise (generalized Langevin equation \cite{Mori:1965}) based on our framework.
The effect of hydrodynamic interaction between particles requires careful consideration. 
Because the hydrodynamic effect may be crucial for applying our theory to macromolecules \cite{Doi:1987} such as biomolecular machinery, we should examine this problem in greater detail.

Since our theory is based on the assumption of the separation of time scales, it cannot be applied to cases in which the separation of time scales is not distinct.
For example, our theory currently does not cover an atomic level description of traditional nonequilibrium systems such as shear flow systems, heat conduction systems and electric conduction systems. Even in such a case, we believe that we can obtain some information on a system by quantifying a degree of fluctuation-dissipation violation. 
More research is required in this regard.

\begin{acknowledgments}
The authors acknowledge K. Hayashi, N. Nakagawa, H. Tasaki, H. Teramoto, and A. Yoshimori for discussions on several issues in this paper.
This study was supported by a grant from the Ministry of Education, Science, Sports and Culture of Japan, No. 16540337 and Research Fellowships for Young Scientists from the Japan Society for the Promotion of Science, No. 05494.
\end{acknowledgments}

\appendix

\section{Fluctuation-dissipation theorem} \label{a.FDT}
In this appendix, we derive the fluctuation-dissipation theorem for the case of equilibrium.
Although the following argument is applicable to the overdamped system, it can be extended to the underdamped case without much difficulty; both cases yield the same result.

Let $\Vec x \equiv (x_0,\cdots,x_{N-1})$ be  a set of dynamical variables under study.
Let the evolution equation of $x_i$ be expressed as
\begin{equation}
\gamma_i \dot x_i(t) = -\pder{U(\Vec x(t) )}{x_i}+\ve 
\fp_i(t) +\xi_i(t), 
\label{eq:lan} 
\end{equation}
where $\ve \fp_i (t)$ denotes a small perturbation force and 
$\xi_i(t)$ represents Gaussian white noise that satisfies
\begin{equation}
\bra \xi_i(t)\xi_j(t')\ket =2\gamma_i T\delta_{ij} \delta(t-t').
\label{noise}
\end{equation}
The initial condition of $x_i(t)$ is set at $t=-\infty$.


In this model, the response function $R_{ij}(t)$ is defined as
\begin{equation}
 \bra \dot x_i(t)\ket_\ve = \ve \int_0^\infty \sum_{j=0}^{N-1}
R_{ij}(s)\fp_j(t-s) \d s+O(\ve^2).
\label{res}
\end{equation}
It should be noted that $R_{ij}(t)=0$ for $t<0$ due to the causality. 
Since $R_{ij}(t)$ does not depend on the selection of \{$\fp_j(t)\}$, 
it can be determined by considering  a special situation in which
$\fp_j(t)=1$ for $t \ge 0$ and $\fp_j(t)=0$ for $t <0$, only for
a specific value of $j$. In this case, Eq.~(\ref{res}) becomes
\begin{equation}
\bra \dot x_i(t)\kettr_{\ve, j} = \ve \int_0^t R_{ij}(s) \d s ,
\label{res2}
\end{equation}
where $\bra \ \kettr_{\ve, j}$ denotes the average for this situation. 

Then, by defining the time-correlation function of velocity as
\begin{equation}
C_{ij}(t) \equiv \bra \dot x_i(t) \circ \dot x_j(0) \ket_0,
\label{cor}
\end{equation}
the fluctuation-response relation implies
\begin{equation}
C_{ij}(t)=T(R_{ij}(t)+R_{ji}(-t)).
\label{FDT}
\end{equation}
In the following, first, Eq.~(\ref{FDT}) is proved by 
focusing on the single component case ($N=1$). Next, a conventional 
derivation  without mathematical rigor is briefly discussed,
which might be useful to argue physical problems.
Finally, the generalization of the proof to the multi-component case is explained briefly, since it is straightforward.

\subsection{Discretized form}\label{sec:a-fdt:dmodel}

As observed in Eqs.~(\ref{eq:lan}) and (\ref{noise}),
$C(0)$ and $R(0)$ are divergent. Therefore, in order to state the 
theorem described in Eq.~(\ref{FDT}) without ambiguity, we
investigate the discretized form of Eq.~(\ref{eq:lan}):
\begin{eqnarray}
\gamma(x^{k+1}-x^k)
&=&-\der{U(x^k)}{x^k}\Delta t +\sqrt{2\gamma T}\Delta W^k+\ve f^k \Delta t \nonumber \\
& & +O\left( (\Delta t)^{3/2}\right),
\label{eq:dlan}
\end{eqnarray}
where  
$\Delta t$ represents the time interval of the discretization; we set $\gamma = \gamma_0$, $x^k=x_0(k\Delta t)$, $\Delta W^k = W((k+1)\Delta t) - W(k \Delta t)$, and $f^k = \fp(k \Delta t)$.
Further, $\Delta W^k$ obeys the Gaussian  distribution with
\begin{equation}
\bra \Delta W^k \Delta W^l \ket = \delta_{kl} \Delta t.
\label{Wnoise}
\end{equation}
It should be noted that  in  Eq.~(\ref{eq:dlan}), the estimation $\Delta W^k =O( (\Delta t)^{1/2})$ 
is assumed, which is expected from Eq.~(\ref{Wnoise}).


In this discretized model represented by Eq.~(\ref{eq:dlan}), 
the time correlation function $C^k$ is defined as 
\begin{equation}
C^k = \bra \frac{x^{k+1}-x^k}{\Delta t} \frac{x^{1}-x^{0}}{\Delta t} \ket_0. 
\label{cor-dis}
\end{equation}
Similarly, by discretizing Eq.~(\ref{res2}), the response function $R^k$ is defined as 
\begin{equation}
\bra \frac{x^{k}-x^{k-1}}{\Delta t} 
\kettr_{\ve, 0} = \ve \sum_{l=0}^{k-1} \Delta t  R^{l},
\label{res-dis}
\end{equation}
where $f^k=1$ for $k \ge 0$ is assumed.  
Therefore, the fluctuation-response relation in Eq.~(\ref{FDT}) should be
regarded  as the continuum form of the relation
\begin{equation}
C^k =T (R^k +R^{-k})
\label{FDT:dis}
\end{equation}
in the limit $\Delta t \to 0$, 
$k \to \infty$ for fixed $k\Delta t $. 
Moreover, $R^k=0$ for $k <0$,  $C^0=2T R^0$ for $k = 0$, 
and $C^k=T R^k$ for $k >0$. 


\subsection{Proof of Eq.~(\ref{FDT:dis})}\label{sec:a-fdt:proof}

First, from Eqs.~(\ref{cor-dis}) and (\ref{res-dis}), Eq.~(\ref{FDT:dis}) is explicitly written as 
\begin{eqnarray}
\bra \frac{x^{k}-x^{k-1}}{\Delta t} \kettr_{\ve, 0} 
&=& \ve \beta  \Delta t \sum_{l=0}^{k-1} \theta^l
\bra \frac{x^{l+1}-x^l}{\Delta t} \frac{x^{1}-x^{0}}{\Delta t} \ket_0 \nonumber \\
& &+O(\ve^2), 
\label{dis-re}
\end{eqnarray}
where $\theta^k=1$ for $k \ge 1$ and $\theta^0=1/2$.
The following is a proof of this expression.

The transition probability $P_\ve(x^k \to x^{k+1})$ 
from $x^k$ to $x^{k+1}$ (for $k \ge 0$) is determined from
\begin{equation}
P_\ve(x^k \to x^{k+1}) \d x^{k+1}= \d (\Delta W^k) 
\sqrt{\frac{1}{2\pi \Delta t}}
\e^{-\frac{(\Delta W^k)^2}{2 \Delta t}}.
\end{equation}
By using Eq.~(\ref{eq:dlan}), this transition probability is calculated as
\begin{eqnarray}
\lefteqn{ P_\ve(x^k \to x^{k+1})= } \\
& & \sqrt{\frac{T}{\pi\gamma \Delta t}}
\e^{-\frac{\beta}{4\gamma  \Delta t}\left[\gamma(x^{k+1}-x^k) 
+\left(\der{U(x^k)}{x^k} -\ve \right) \Delta t+O\left( (\Delta t)^{3/2}\right)\right]^2}. \nonumber
\end{eqnarray}
Based on the estimation
\begin{eqnarray}
\lefteqn{U(x^{k+1})-U(x^k)=} \\
& & \frac{1}{2}
\left(\der{U(x^k)}{x^k}+\der{U(x^{k+1})}{x^{k+1}} \right)
(x^{k+1}-x^k) +O\left((\Delta t)^{3/2}\right), \nonumber
\label{est}
\end{eqnarray}
it is confirmed that 
\begin{eqnarray}
\lefteqn{\frac{P_\ve(x^k \to x^{k+1})}{P_\ve(x^{k+1} \to x^{k})} =}\nonumber \\
& & \e^{-\beta \left[ U(x^{k+1})-U(x^k)-\ve (x^{k+1}-x^k) \right] +O\left((\Delta t)^{3/2}\right)}.
\label{ldb}
\end{eqnarray}

Since an initial condition is imposed at $t=-\infty$, the 
$(k+1)$-time probability distribution function at time $t=t^l$ for
$l=0,1,\cdots,k$ is expressed as
\begin{equation}
P_\ve(x^0,\cdots,x^k)=\pc(x^0)\prod_{l=0}^{k-1} P_\ve(x^l\to x^{l+1}),
\label{defP}
\end{equation}
where $\pc(x)$ is the canonical distribution 
\begin{equation}
\pc(x)=\frac{1}{Z}\e^{-\beta U(x)}.
\end{equation}
The key identity to derive Eq.~(\ref{FDT:dis}) is 
\begin{equation}
\frac{P_\ve(x^0,\cdots,x^k)}{P_\ve(x^k,\cdots,x^0)}
=\e^{\ve \beta (x^k-x^0)+O( \sum_{l=0}^k (\Delta t)^{3/2})}, 
\label{key}
\end{equation}
which is easily obtained from  Eq.~(\ref{defP}) and 
Eq.~(\ref{ldb}). Since $\sum_{l=0}^k (\Delta t)^{3/2} \to 0$
in the limit $\Delta t \to 0$, $k \to \infty$ for fixed $k \Delta t $, 
the term $O( \sum_{l=0}^k (\Delta t)^{3/2}) $ in Eq.~(\ref{key}) can be 
neglected. 

The left-hand side of Eq.~(\ref{dis-re}) is evaluated as follows.
First, using the key identity Eq.~(\ref{key}), we calculate
\begin{eqnarray}
\lefteqn{
\int \prod_{l = 0}^{k} \d x^lP_\ve(x^0,\cdots,x^k)(x^{k}-x^{k-1})} \hspace{0mm}
\nonumber \\
&=&
\int \prod_{l = 0}^{k} \d x^lP_\ve(x^k,\cdots,x^0)(x^{k}-x^{k-1})
\e^{\ve \beta (x^k-x^0)}
\nonumber \\
&=&
-\int \prod_{l = 0}^{k} \d x^lP_\ve(x^0,\cdots,x^k)(x^{1}-x^{0})
\e^{-\ve \beta (x^k-x^0)}
\nonumber \\
&=& 
-\int \prod_{l = 0}^{k} \d x^lP_\ve(x^0,\cdots,x^k)(x^{1}-x^{0})
(1-\ve \beta (x^k-x^0)) \nonumber \\
& & +O(\ve^2).
\label{cal}
\end{eqnarray}
By setting $k=1$ in this expression, we obtain
\begin{eqnarray}
\lefteqn{\int \d x^0 \d x^1 P_\ve(x^0,x^1)(x^{1}-x^{0}) } \\
&=& \frac{1}{2}\ve \beta \int \d x^0 \d x^1 P_0(x^0,x^1)(x^{1}-x^{0})^2
+O(\ve^2). \nonumber
\end{eqnarray}
Next, Eq.~(\ref{cal}) can be rewritten as
\begin{eqnarray}
\lefteqn{ \int \prod_{l = 0}^k \d x^l P_\ve(x^0,\cdots,x^k)(x^{k}-x^{k-1}) =} \nonumber \\
& &
\ve \beta \int \prod_{l = 0}^{k} \d x^l 
P_0(x^0,\cdots,x^k) \sum_{l'=0}^{k-1} (x^{1}-x^{0})(x^{l'+1}-x^{l'})\theta^l
\nonumber \\
& & +O(\ve^2).
\label{dis0}
\end{eqnarray}
By dividing both sides by $\Delta t$, Eq.~(\ref{dis-re}) is obtained.

\paragraph*{Detailed balance}

It should be noted that  Eq.~(\ref{key}) is essential to derive the fluctuation-
dissipation relation. The condition in Eq.~(\ref{key}) implies
a time-reversal symmetry expressed as
\begin{equation}
P_0(x^0,\cdots,x^N)=P_0(x^N,\cdots,x^0);
\label{DB}
\end{equation}
it is referred to as the detailed-balance condition. In general, if this condition does not hold for a system without 
a probe force ($\ve=0$), the fluctuation-response relation
cannot be derived. In fact, for models studied in this
paper, this relation  does not hold in
nonequilibrium steady states because the detailed balance 
is violated.
 
\subsection{Conventional derivation}\label{sec:a-fdt:conv} 

When the mathematical rigor is not seriously cared, the 
fluctuation-response relation of Eq.~(\ref{FDT}) can be quickly
derived by employing the path integral representation:
\begin{eqnarray}
\lefteqn{T([x])=} \\ 
& & K \exp\left[-\frac{\beta}{4\gamma}\int_0^\tau 
\left(\gamma \dot x(t) +\der{U(x(t))}{x}-\ve\fp (t)\right)^2  \d t \right], \nonumber
\end{eqnarray}
where $T([x])$ denotes the probability density of trajectory $x(t)$, 
$0 < t \le \tau$, provided that $x(0)$ is given; $K$ is a normalization
constant. By denoting the time reversed trajectory of $x(t)$ by 
$\tilde x(t)=x(\tau-t)$, we obtain
\begin{equation}
\frac{T([x])}{T([\tilde x])}=\exp\left[-\beta 
\int_0^\tau \dot x \left( \der{U(x(t))}{x}-\ve\fp(t) \right) \d t \right].
\end{equation}
Using this equation, the following identity is obtained for an arbitrary quantity $A([x])$:
\begin{eqnarray}
\bra A \ket &=& \int {\cal D}[x]\pc(x(0))T([x])A([x]) \nonumber \\
&=& \int {\cal D}[x]\pc(x(0))T([x]) \e^{-\beta \ve\int_0^\tau  \d t \dot x(t)\fp(t)}
\tilde A([x]) \nonumber  \\
&=& \bra  \e^{-\beta \ve\int_0^\tau  \d t \dot x(t)\fp(t)} \tilde A \ket,
\end{eqnarray}
where $\tilde A ([x])\equiv A([\tilde x])$.
By setting $A([x])=\dot x(t)|_{t=\tau}$, we obtain the equality 
$C(t)=T R(t)$ for $t>0$. This corresponds to Eq.~(\ref{FDT}) 
for $N=1$. 

\subsection{Multi-component case}\label{sec:a-fdt:multi} 

By introducing discretized variables 
$\Vec x^k = \{ x_i^k \} \equiv \{ x_i(k\Delta t) \}$, 
$0 \le i \le  N-1$, the proof described in Appendixes \ref{sec:a-fdt:dmodel}
and \ref{sec:a-fdt:proof} can be generalized to the multi-component 
case.  When only the $j^\mathrm{th}$ probe force is applied from $t=0$ ($\fp_j(t)=1$ for $t \ge 0$ and $\fp_j(t)=0$ for $t <0$),
Eq.~(\ref{key}) becomes
\begin{equation}
\frac{P_{\ve, j} (\Vec x^0 ,\cdots,\Vec x^k)}
{P_{\ve, j} (\Vec x^k ,\cdots,\Vec x^0)}
=\e^{\ve \beta (x_j^k-x_j^0)+O( \sum_{l=0}^k (\Delta t)^{3/2})},
\label{key-multi}
\end{equation}
where $P_{\ve, j} (\Vec x^0 ,\cdots,\Vec x^k)$ is the joint probability distribution in the presence of the above mentioned probe force.
Then, the following identity is obtained:
\begin{eqnarray}
\lefteqn{
\int \prod_{l = 0}^k \d \Vec x^l
P_{\ve, j} (\Vec x^0,\cdots,\Vec x^k)(x_{i}^k-x_i^{k-1}) =} \hspace{0mm}
\nonumber \\
& &
\ve \beta \int \prod_{l = 0}^k \d \Vec x^l
P_0(\Vec x^0,\cdots,\Vec x^k) 
\sum_{l=0}^{k-1} (x^{1}_i-x^{0}_i)(x^{l+1}_j-x^l_j)\theta^l \nonumber \\
& & +O(\ve^2).
\label{dis0-multi}
\end{eqnarray}

By defining
\begin{equation}
C^k_{ij}=\bra \frac{x^{k+1}_i-x^k_i}{\Delta t}  \frac{x^{1}_j-x^{0}_j}{\Delta t}
\ket_0, 
\label{cor-dis:multi}
\end{equation}
and 
\begin{equation}
\bra \frac{x^{k}_i-x^{k-1}_i}{\Delta t} 
\kettr_{\ve, j} = \ve \sum_{l=0}^{k-1} \Delta t  R^l_{ij},
\label{res-dis:multi}
\end{equation}
Eq.~(\ref{dis0-multi}) leads to
\begin{equation}
C^k_{ji}=T R^k_{ij} \theta^k
\end{equation}
for $k \ge 0$. Since $C^k_{ij}=C^{-k}_{ji}$, this relation is
written as
\begin{equation}
C^k_{ij}=T (R^k_{ij} +R^{-k}_{ji})
\end{equation}
for all $k$. By taking the limit $\Delta t \to 0$, we obtain Eq.~(\ref{FDT}).

A matrix representation can be used in the multi-component case.
Let ${\cal C}(t)$ and ${\cal R}(t)$ be the matrices 
whose $(i,j)$ components are $C_{ij}(t)$ and $R_{ij}(t)$, respectively.
Using these matrices, Eq.~(\ref{FDT}) is expressed as
\begin{equation}
{\cal C}(t)=T ({\cal R}(t)+{\cal R}^\dagger(-t) ),
\end{equation}
where the symbol ${}^\dagger$ represents the transpose of the matrix. 
From this equation, the following relations are derived with regard to the symmetric and antisymmetric parts
of the matrices:
\begin{eqnarray}
{\cal C}^\mathrm{s}(t) &=& 
T ({\cal R}^\mathrm{s}(t)+{\cal R}^\mathrm{s}(-t) ), \\
{\cal C}^\mathrm{a}(t) &=& 
T ({\cal R}^\mathrm{a}(t)-{\cal R}^\mathrm{a}(-t) ), 
\end{eqnarray}
where the symbols $^\mathrm{s}$ and $^\mathrm{a}$ denote 
the symmetric and antisymmetric parts, respectively.

\section{Quick derivation of the equality} \label{a.quick}

Based on the path-integral argument presented in Appendix \ref{sec:a-fdt:conv}, we can simplify the derivation of Eq. (\ref{e.th}).
In this appendix, we do not consider the mathematical rigor, although the following argument can be made more precise by employing the discretization argument, as shown in Appendix \ref{sec:a-fdt:dmodel}.
For simplicity, we consider the case with a single degree of freedom without the inertia term. The generalization of the argument to multivariable and underdamped cases is straightforward.

We analyze the Langevin equation
\begin{equation}
\gamma \dot x(t) = F\left( x(t) \right) + \xi(t) + \ve \fp (t),
\label{e.singlemodel}
\end{equation}
where $\xi(t)$ denotes the zero-mean white Gaussian noise that satisfies
\begin{equation}
\bra \xi(t)\xi(t')\ket=2\gamma T\delta(t-t').
\end{equation}
The probability of a trajectory $[x] = \{ x(t) \}$, $t_0 \le t \le t_1$ for $x = x_0$ at $t = t_0$ is expressed as
\begin{equation}
{\cal D}[x] P(x_0 | [x]) = {\cal D}[\xi] K \e^{-\frac{\beta}{4\gamma} \int_{t_0}^{t_1} \left[ \gamma \dot x (t) - F(x(t)) - \ve \fp (t) \right]^2 \d t},
\label{e.path}
\end{equation}
where $K$ is a normalization constant.
By using this probability, we obtain
\begin{equation}
\bra \dot x(t) \ket_\ve = \int {\cal D}[x] P(x_0 | [x]) \dot x(t),
\label{e.dotx}
\end{equation}
for $t_0 \le t \le t_1$. From Eq.~(\ref{e.dotx}), 
\begin{equation}
\frac{\delta \bra \dot x(t) \ket_\ve}{\delta (\ve \fp(s))} = \frac{\beta}{2\gamma} \bra \dot x(t) \circ \left[ \gamma \dot x(s) - F\left( x(s) \right) \right] \ket_\ve.
\label{e.dxdf}
\end{equation}
Based on the definition of the response function, the left-hand side of Eq.~(\ref{e.dxdf}) is identified as $R(t-s)$ in the limit $\ve \to 0$. Therefore, in this limit,
\begin{equation}
R(t-s) = \frac{\beta}{2} \left[ \vs^2 + C(t-s) \right] - \frac{\beta}{2\gamma} \bra \dot x(t) \circ F\left( x(s) \right) \ket_0.
\label{e.vio}
\end{equation}
By exchanging $t$ and $s$ in Eq.~(\ref{e.vio}) and adding the resulting 
expression to Eq.~(\ref{e.vio}), we obtain
\begin{eqnarray}
\lefteqn{R(t-s) + R(s-t) = \beta \left[ \vs^2 + C(t-s) \right]} \hspace{5mm} \nonumber \\
& & - \frac{\beta}{2 \gamma} \bra \dot x(t) \circ F\left( x(s) \right) + \dot x(s) \circ F\left( x(t) \right) \ket_0.
\label{e.vio2}
\end{eqnarray}
Since the last term on the right-hand side of Eq.~(\ref{e.vio2}) becomes $\bra J \ket_0$ when $t = s$, Eq.~(\ref{e.th}) is obtained.

\section{Derivation of Lemma} \label{a.lemma}

For an arbitrary non-anticipating function $A(t)$ \cite{Gardiner:2004}, the following relation holds.
\begin{enumerate}
\item When $t < t_N$,
\begin{eqnarray}
\lefteqn{\bra \left( \int_{t_0}^{t_N} A(t_N -s) \cdot \d W(s) \right)\circ \d W(t) \ket_0 } \hspace{35mm}\nonumber \\
&=& \bra A(t_N - t) \ket_0 \d t.
\label{e.lemma1}
\end{eqnarray}
\item When $t = t_N$,
\begin{eqnarray}
\lefteqn{\bra \left( \int_{t_0}^{t_N} A(t_N -s) \cdot \d W(s) \right)\circ \d W(t) \ket_0 } \hspace{35mm} \nonumber \\
&=& \frac{1}{2} \bra A(0+) \ket_0 \d t.
\label{e.lemma2}
\end{eqnarray}
\item When $t > t_N$,
\begin{equation}
\bra \left( \int_{t_0}^{t_N} A(t_N -s) \cdot \d W(s) \right)\circ \d W(t) \ket_0 = 0.
\label{e.lemma3}
\end{equation}
\end{enumerate}

\subsection*{Proof}
We discretize the time interval $(t_0, t_N)$ as $t_0 < t_1 < t_2 < \cdots < t_{N-1} < t_N$.
In the following, the symbol $\simeq$ is used to imply equality in the limit of $N \to \infty$.
Further, we use the notation, $\Delta W^k \equiv W(t_{k+1}) - W(t_k)$.

\noindent 1. When $t < t_N$, 
$t_k$ can be considered such that $t = t_k$.
By discretizing the left-hand side of Eq.~(\ref{e.lemma1}), we obtain
\begin{widetext}
\begin{eqnarray}
\bra \left[ \int_{t_0}^{t_N} A(t_N - s) \cdot \d W(s) \right] \circ \d W(t) \ket &=& \bra \left[ \int_{t_0}^{t_k} A(t_N - s) \cdot \d W(s) + \int_{t_k}^{t_N} A(t_N - s) \cdot \d W(s) \right] \circ \d W(t_k) \ket_0 \nonumber \\
&\simeq& \bra \frac{1}{2} \left[ \int_{t_0}^{t_{k+1}} A(t_N - s) \cdot \d W(s) + \int_{t_0}^{t_k} A(t_N - s) \cdot \d W(s) \right] \Delta W^k \ket_0 \nonumber \\
& & + \bra \frac{1}{2} \left[ \int_{t_{k+1}}^{t_N} A(t_N - s) \cdot \d W(s) + \int_{t_k}^{t_N} A(t_N - s) \cdot \d W(s) \right] \Delta W^k \ket_0 \nonumber \\
&\simeq& \bra \frac{1}{2} \left[ \sum_{l=0}^k A(t_N - t_l) \Delta W^l + \sum_{l=0}^{k-1} A(t_N - t_l) \Delta W^l \right] \Delta W^k \ket_0 \nonumber \\
& & + \bra \frac{1}{2} \left[ \sum_{l=k+1}^{N-1} A(t_N - t_l) \Delta W^l + \sum_{l=k}^{N-1} A(t_N - t_l) \Delta W^l \right] \Delta W^k \ket_0.
\label{e.lp1}
\end{eqnarray}
\end{widetext}
Since $A(t)$ is a non-anticipating function and that $\Delta W^k$ is independent of $\Delta W^l$ when $k \neq l$, Eq.~(\ref{e.lp1}) can be written as
\begin{eqnarray}
\lefteqn{
\bra \left[ \int_{t_0}^{t_N} A(t_N - s) \cdot \d W(s) \right] \circ \d W(t) \ket_0 
} \hspace{0mm} \nonumber \\
&\simeq& \bra \frac{1}{2} A(t_N - t_k) \Delta W^k \Delta W^k + \frac{1}{2} A(t_N - t_k) \Delta W^k \Delta W^k \ket_0 \nonumber \\ 
&=& \bra A(t_N - t_k) \ket_0 \bra (\Delta W^k )^2 \ket_0 \nonumber \\
&=& \bra A(t_N - t_k) \ket_0 (t_{k+1} - t_j).
\label{e.lp2}
\end{eqnarray}
By considering the limit $N \to \infty$, we obtain Eq.~(\ref{e.lemma1}).

\noindent 2. For $t = t_N$, 
we define $t_{N+1}$ such that $\d W(t_N) \simeq W(t_{N+1}) - W(t_N)$.
By discretizing the left-hand side of Eq.~(\ref{e.lemma2}), we obtain
\begin{widetext}
\begin{eqnarray}
\bra \left[ \int_{t_0}^{t_N} A(t_N - s) \cdot \d W(s) \right] \circ \d W(t) \ket
&\simeq& \bra \frac{1}{2} \left[ \int_{t_0}^{t_{N+1}} A(t_{N+1} - s) \cdot \d W(s) + \int_{t_0}^{t_N} A(t_N - s) \cdot \d W(s) \right] \Delta W^N \ket_0 \nonumber \\
&\simeq& \bra \frac{1}{2} \left[ \sum_{k=0}^N A(t_{N+1} - t_k) \Delta W^k + \sum_{k=0}^{N-1} A(t_N - t_k) \Delta W^k \right] \Delta W^N \ket_0.
\label{e.lp3}
\end{eqnarray}
\end{widetext}
Since $A(t)$ is a non-anticipating function and that $\Delta W^k$ is independent of $\Delta W^l$ when $k \neq l$, Eq.~(\ref{e.lp1}) can be written as
\begin{eqnarray}
\lefteqn{
\bra \left[ \int_{t_0}^{t_N} A(t_N -s) \cdot \d W(s) \right] \circ \d W(t) \ket_0 } \hspace{10mm} \nonumber \\
&\simeq& \frac{1}{2} \bra A(t_{N+1} - t_N) \ket_0 \bra (\Delta W^N)^2 \ket_0 \nonumber \\
&=&\frac{1}{2} \bra A(t_{N+1} - t_N) \ket_0 (t_{N+1} - t_N).
\end{eqnarray}
By considering the limit $N \to \infty$, we obtain Eq.~(\ref{e.lemma2}).

\noindent 3. For $t > t_N$, $\d W(t)$ is independent of $\d W(s)$ for $s \le t_N < t$; hence, Eq.~(\ref{e.lemma3}) is obtained immediately.

\bibliography{biblio.bib}

\end{document}